\documentclass[acmsmall,screen]{acmart}

\usepackage{multirow}
\usepackage{longtable}
\usepackage{tabularx}
\usepackage{array}
\usepackage[ruled,vlined,linesnumbered]{algorithm2e}
\usepackage{tcolorbox}
\usepackage{pifont}
\usepackage{wasysym}
\usepackage{graphicx}
\usepackage{tikz}     
\usepackage{mdframed}   
\usepackage{caption} 
\usepackage{subcaption} 
\usepackage{enumitem}
\usepackage{siunitx}
\usepackage{booktabs}
\usepackage{colortbl}
\usepackage{xcolor}
\usepackage{nicematrix}

\newcommand{\tool}{{\textsc{SimADFuzz} }}

\newcommand{\ie}{{\it i.e., }}
\newcommand{\eg}{{\it e.g., }}

\newcommand{\cf}{{\it cf. }}

\AtBeginDocument{%
  }
 
\setcopyright{acmlicensed}
\copyrightyear{2024}
\acmYear{2024}
\acmDOI{XXXXXXX.XXXXXXX}

\begin{document}

\title{SimADFuzz: Simulation-Feedback Fuzz Testing for Autonomous Driving Systems}

\author{Huiwen Yang}
\email{yhw\_yagol@nuaa.edu.cn}
\affiliation{
	\institution{Nanjing University of Aeronautics and Astronautics}
	\city{Nanjing}
	\country{China}
}

\author{Yu Zhou}
\authornote{Corresponding author.}
\email{zhouyu@nuaa.edu.cn}
\affiliation{
	\institution{Nanjing University of Aeronautics and Astronautics}
	\city{Nanjing}
	\country{China}
}

\author{Taolue Chen}
\authornotemark[1]
\email{t.chen@bbk.ac.uk}
\affiliation{
	\institution{Birkbeck, University of London}
	\city{London}
	\country{United Kingdom}
}

\renewcommand{\shortauthors}{Yang et al.}

\begin{abstract}
	Autonomous driving systems (ADS) have achieved remarkable progress in recent years. However, ensuring their safety and reliability remains a critical challenge due to the complexity and uncertainty of driving scenarios. In this paper, we focus on simulation testing for ADS, where generating diverse and effective testing scenarios is a central task.
	Existing fuzz testing methods face limitations, such as overlooking the temporal and spatial dynamics of scenarios and failing to leverage simulation feedback (\eg speed, acceleration and heading) to guide scenario selection and mutation.
	To address these issues, we propose \tool, a novel framework designed to generate high-quality scenarios that reveal violations in ADS behavior. Specifically, \tool employs violation prediction models, which evaluate the likelihood of ADS violations, to optimize scenario selection.
	Moreover, \tool proposes distance-guided mutation strategies to enhance interactions among vehicles in offspring scenarios, thereby triggering more edge-case behaviors of vehicles.
	Comprehensive experiments demonstrate that \tool outperforms state-of-the-art fuzzers by identifying 32 more unique violations, including 4 reproducible cases of vehicle-vehicle and vehicle-pedestrian collisions. These results demonstrate \tool’s effectiveness in enhancing the robustness and safety of autonomous driving systems.
\end{abstract}

\ccsdesc[500]{Software and its engineering~Software testing and debugging}

\keywords{Autonomous Driving Systems, Fuzz Testing, Simulation-based Testing}

\maketitle

\section{Introduction}
Autonomous driving systems (ADS), such as Apollo\footnote{Apollo, \url{https://github.com/ApolloAuto/apollo}} and Autoware\footnote{Autoware, \url{https://github.com/autowarefoundation/autoware}}, have made significant progress in recent years. Various technologies have been developed to enable vehicles to operate autonomously without human intervention~\cite{chan2017advancements,Chen_2015_ICCV}. However, ensuring the safety and reliability of these systems remains a critical challenge~\cite{adsbugs}. As of March 2024, the California Department of Motor Vehicles reported 695 traffic accidents involving autonomous vehicles\footnote{DMV Autonomous Vehicle Collision Reports, \url{https://www.dmv.ca.gov/portal/vehicle-industry-services/autonomous-vehicles/autonomous-vehicle-collision-reports/}}, including 133 collisions in 2023. These incidents underscore the urgent need for comprehensive and effective testing of ADS before deployment.

Recent research has focused on leveraging various technologies to test the performance and reliability of ADS. For instance, adversarial attacks, widely used in computer vision, have been applied to test the robustness of ADS perception modules by exposing vulnerabilities in object detection and classification~\cite{DeepBillboard,WangS0CZZ0M22}. Similarly, software testing techniques such as search-based testing~\cite{9724804} and fuzz testing~\cite{AV-Fuzzer,AutoFuzz,DriveFuzz,Doppel} have shown great potential in identifying defects and vulnerabilities in ADS. While these methods have achieved significant results, challenges remain in ensuring comprehensive coverage and scalability for real-world scenarios.

Simulation-based testing has emerged as a widely adopted method for evaluating ADS due to its efficiency and cost-effectiveness compared to real-road testing~\cite{9706219}. By generating diverse and realistic driving scenarios, simulation-based testing can evaluate ADS under various conditions, identifying potential violations such as collisions, unsafe lane changes, and traffic rule violations. This approach is indispensable for uncovering safety issues and improving the reliability of ADS.

The quality of simulation scenarios is critical to the effectiveness of simulation-based testing for ADS~\cite{dai2024sctrans}. Various scenario generation methods, such as DriveFuzz~\cite{DriveFuzz} and Doppel~\cite{Doppel}, have been proposed. In general, these methods leverage genetic algorithms to generate offspring scenarios through selection, crossover, and mutation of parent scenarios. Scenarios are evaluated and prioritized using pre-designed fitness functions, often considering factors such as the behavior of the \emph{ego vehicle}\footnote{The vehicle controlled by ADS is referred to as the ego vehicle, while other vehicles are referred to as \emph{NPC vehicles}.} or minimum distances between vehicles. Random mutation strategies are then applied to generate offspring scenarios. However, these methods face several limitations:

\begin{itemize}
	\item[(1)] \emph{Limitations in scenario fitness evaluation}. Existing fitness functions typically prioritize scenarios based on simple aggregation methods, such as maximum, average, or median values. However, these approaches overlook the sequential nature of driving scenarios, which consist of discrete temporal scenes capturing dynamic interactions and behaviors. Simply aggregating attributes fails to account for these temporal dynamics, potentially leading to suboptimal prioritization.

	\item[(2)] \emph{Limitations in mutation strategies}. Scenarios involve a large number of mutable elements, resulting in an enormous search space. While random mutation is a natural approach, it fails to consider the interactions between mutable elements, such as vehicles and pedestrians. As a result, this strategy may struggle to produce high-quality scenarios that effectively challenge the ADS.
\end{itemize}

In this paper, we propose \tool, a simulation-feedback fuzz testing method for ADS to address the limitations of existing methods. \tool monitors and collects simulation feedback, including the coordinates and physical states of vehicles, during the simulation. Based on genetic algorithms, it generates high-quality testing scenarios. Unlike previous work that primarily uses feedback for scenario selection, \tool innovatively leverages feedback to extract temporal features for scenario fitness evaluation and to design effective mutation strategies.

To address the limitations of scenario fitness evaluation, \tool optimizes scenario selection using model-based fitness evaluation methods. Specifically, it integrates a Transformer~\cite{vaswani2017attention} encoder to capture the continuous and dynamic nature of driving scenarios. The scenarios are embedded into a high-dimensional feature space, enabling the evaluation of scenario priority through violation prediction models that assess the likelihood of triggering unsafe behaviors.

Additionally, to overcome the limitations of mutation strategies, \tool adopts a distance-guided mutation strategy. This strategy dynamically adjusts the mutation probability of NPC vehicles based on their proximity to ego vehicles, increasing the likelihood of interactions. By prioritizing NPC vehicles closer to the ego vehicle for mutation, \tool generates offspring scenarios that are more likely to expose potential safety issues while maintaining scenario diversity.

We conduct extensive experiments on InterFuser, a top-tier agent that secured 2nd place on the CARLA leaderboard\footnote{The 1st place agent, ReasonNet, has not made its code publicly available. Cf.\ CARLA leaderboard, \url{https://leaderboard.carla.org/leaderboard/}}. The results show that \tool detects 10 more unique violations than the baselines, which employs random selection and mutation strategies. Moreover, \tool discovers 32, 27 and 9 more violations than AV-Fuzzer~\cite{AV-Fuzzer}, DriveFuzz~\cite{DriveFuzz} and TM-Fuzzer~\cite{lin2024tm}, respectively. In the end, \tool totally found 4 collisions triggered by InterFuser, which shows the effectiveness in detecting ADS violations.


The main contributions of this paper are summarized as follows:

\begin{enumerate}

	\item We propose a novel fuzz testing method for ADS, named \tool, which leverages simulation feedback to generate high-quality scenarios. \tool effectively discovers violations in ADS by dynamically analyzing vehicle states and interactions during simulation.

	\item We develop a model-based scenario fitness evaluation approach. By utilizing violation prediction models and incorporating a Transformer encoder, \tool captures the temporal features of driving scenarios, enabling more accurate prioritization of high-risk scenarios.

	\item We introduce distance-guided mutation strategies that mutate NPC vehicles based on their proximity to ego vehicles. This approach increases the likelihood of interactions, generating diverse and challenging scenarios that expose potential safety issues in ADS.

	\item We conduct extensive experiments to evaluate the effectiveness of \tool. Results demonstrate that \tool detects significantly more violations compared to state-of-the-art methods, including collisions and lane invasion. To facilitate reproducibility and further research, we release the implementation publicly.
\end{enumerate}

\noindent\emph{Structure.} The rest of this paper is organized as follows. Section~\ref{preliminary} introduces the preliminaries of simulation-based testing and genetic algorithms. Section~\ref{sect:example} illustrates the limitations of existing methods through a motivating example. Section~\ref{sect:approach} details the design and implementation of \tool. Section~\ref{sect:exp} presents the experimental results and discusses potential threats to validity. Section~\ref{sect:related} reviews related work in the field. Finally, Section~\ref{sect:conc} concludes the paper and proposes directions for future research.
\section{Preliminaries}
\label{preliminary}

Simulation-based testing is a widely adopted approach to evaluate the performance and reliability of ADS~\cite{10.1145/3540250.3549111,10.1145/3338906.3338942,6957844}. This method uses virtual scenarios as structured test cases to execute and validate target systems. Genetic algorithms~\cite{8863940,10.1145/3133956.3134020}, inspired by natural selection, are commonly employed in fuzz testing to generate diverse and high-quality test cases. Several fuzz testing methods, such as AV-Fuzzer~\cite{AV-Fuzzer}, DriveFuzz~\cite{DriveFuzz}, Doppel~\cite{Doppel}, and TM-Fuzzer~\cite{lin2024tm}, have been proposed for simulation-based ADS testing.

\begin{figure}[h]
	\centering
	\includegraphics[width=0.8\linewidth]{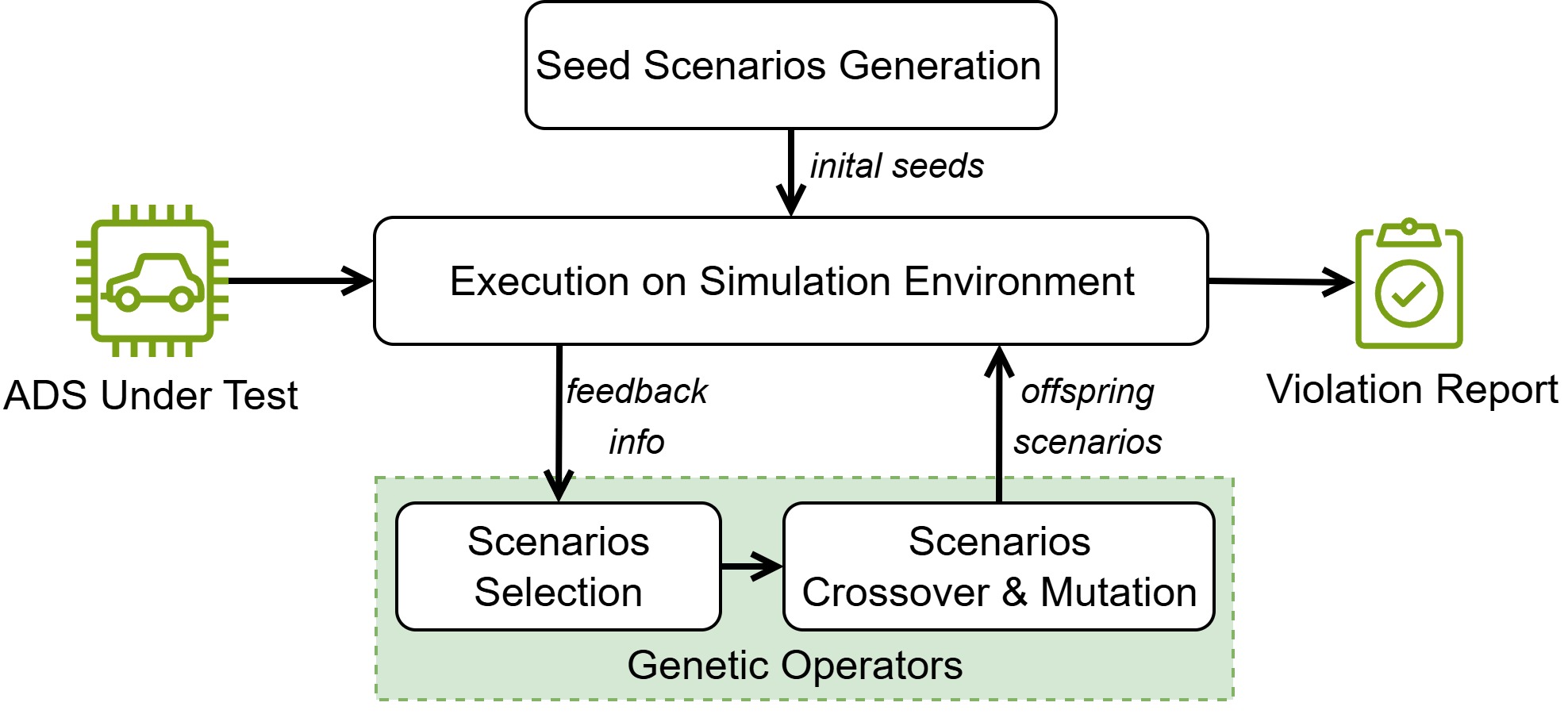}
	\caption{Simulation-based Fuzz Testing Framework for ADS}
	\label{fig:fuzz}
\end{figure}

Figure~\ref{fig:fuzz} illustrates the framework of simulation-based fuzz testing for ADS. This framework takes the ADS under test as input and produces violation reports as output. It consists of three main components: seed scenario generation, simulation, and genetic operators. First, seed scenarios are generated and executed in the simulation environment. During the execution, feedback information such as vehicle coordinates and velocities is collected from the simulator to evaluate the fitness of each scenario. Based on these fitness scores, the most promising scenarios are selected. These selected scenarios undergo crossover or mutation to produce offspring scenarios, which are then executed in the simulation environment for subsequent generations. During execution, the simulation environment detects various types of violations, such as collisions, unsafe lane changes, and traffic rule violations. This fuzzing process is repeated until the testing budget (\eg time) is exhausted. At the end of the process, a report summarizing the detected violations is generated to aid in evaluating the ADS's performance and identifying potential issues.

In this framework, \emph{scenario selection}, \emph{crossover}, and \emph{mutation} are the three main genetic operators that define the core of genetic algorithms. For scenario selection, genetic algorithms identify promising scenarios based on several metrics and evaluate their fitness scores using single- or multi-objective search methods. These evaluations are performed using fitness functions, which quantify the likelihood of a scenario causing violations~\cite{alhijawi2023genetic}. Table~\ref{tab:fitness} summarizes the commonly used metrics in simulation-based testing for ADS. As shown in Table~\ref{tab:fitness}, the fitness function evaluates scenarios to estimate their probability of triggering violations. For scenario crossover and mutation, random strategies are widely used. For instance, DriveFuzz~\cite{DriveFuzz} employs random mutation by altering weather conditions (\eg wind, cloud cover, and rain) or modifying pedestrian behaviors without utilizing any crossover operator. Similarly, Doppel~\cite{Doppel} adopts random mutation by adding or removing traffic participants or modifying the starting and destination points of ego vehicles. On the other hand, Doppel's crossover operator swaps ego vehicles between two scenarios when their routes intersect.

\begin{table*}[h]
	\caption{Fitness metrics used in AV-Fuzzer, DriveFuzz, and Doppel}
	\centering
	\begin{tabularx}{\textwidth}{llX}
		\toprule
		\textbf{Method}                                     & \textbf{Factor}              & \textbf{Description}                                                                \\
		\hline
		\multirow{4}{*}{AV-Fuzzer\cite{AV-Fuzzer}} & $d_{safe}$          & The maximum distance without colliding with other actors                   \\
		                                           & $d_{stop}$          & The distance the vehicle will travel before coming to a complete stop      \\ \cline{2-3}
		                                           & Fitness score       & $=d_{safe}-d_{stop}$                                                       \\ \hline
		\multirow{7}{*}{DriveFuzz\cite{DriveFuzz}} & $ha$                & The times of hard acceleration                                             \\
		                                           & $hb$                & The times of hard braking                                                  \\
		                                           & $ht$                & The times of hard turn                                                     \\
		                                           & $os$                & The times of oversteer                                                     \\
		                                           & $us$                & The times of understeer                                                    \\
		                                           & $md$                & The minimum distances from ego vehicle to other actors                     \\ \cline{2-3}
		                                           & Fitness score       & $= - (ha +hb +ht + os + us - 1/md)$                                        \\ \hline
		\multirow{7}{*}{Doppel\cite{Doppel}}       & $f_{min\_distance}$ & The minimum distances from ego vehicle to other actors                     \\
		                                           & $f_{decision}$      & The total number of unique decisions being made by all ego vehicles        \\
		                                           & $f_{conflict}$      & The total number of pairs of actors whose trajectory overlaps with another \\
		                                           & $f_{violation}$     & The total number of violations across all ego vehicles                     \\ \cline{2-3}
		                                           & Fitness score       & Based on NSGA-2                                                            \\ \bottomrule
	\end{tabularx}
	\label{tab:fitness}
\end{table*}

In this paper, we aim to optimize two key genetic operators: \emph{scenario selection} and \emph{mutation}. The details of these optimizations are presented in Section~\ref{sect:approach}.
\section{A Motivating Example} \label{sect:example}

This section presents an example to illustrate the limitations of existing methods and discusses the necessity of introducing model-based fitness evaluation and distance-guided mutation strategies.

\begin{figure}[h]
    \centering
    \includegraphics[width=1\linewidth]{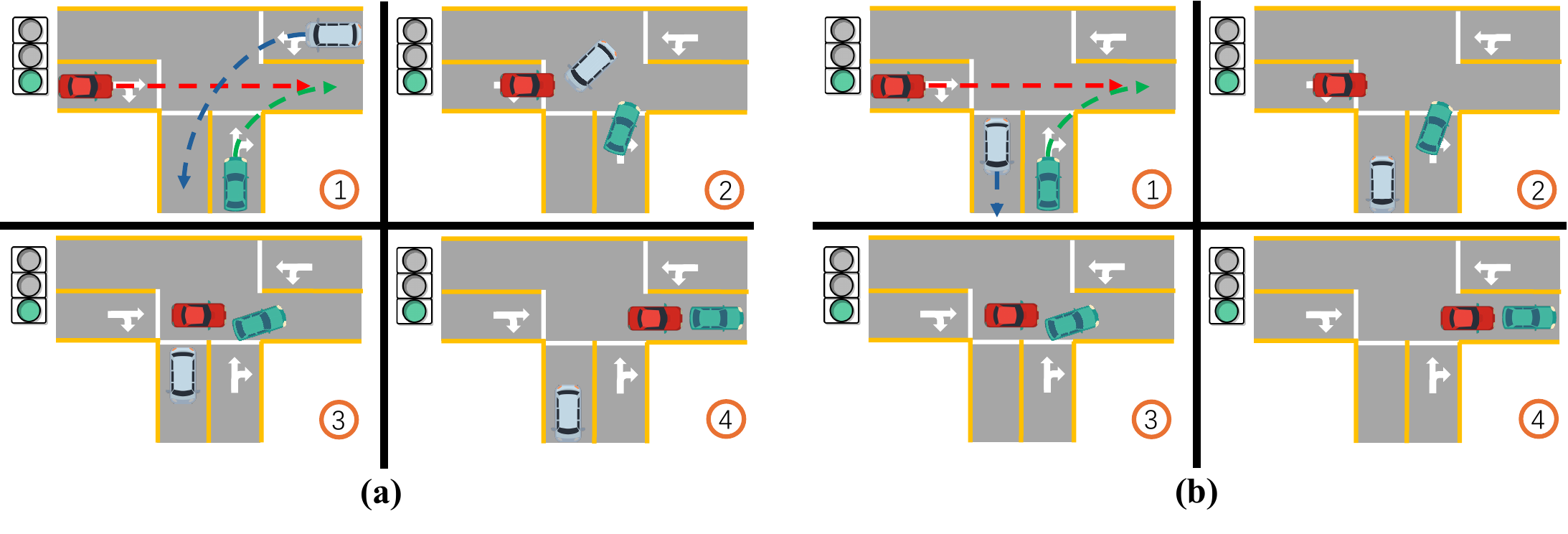}
	\caption{Multi-Vehicle Interaction at a T-junction}
	\label{fig:me}
\end{figure}

As shown in Figure~\ref{fig:me}(a), the scenario involves three vehicles converging at a $T$-junction. The red vehicle intends to proceed straight, the white vehicle is making a left turn, and the green vehicle is turning right. The traffic light for the east-west direction is green, allowing all vehicles to proceed legally.

At time point 2, while the white vehicle is turning left, it encounters the red vehicle travelling straight. The red vehicle brakes to avoid a collision, maintaining a minimum distance of $5$ meters ($md_{t2}=5$) between them. At time point 3, the green vehicle completes its right turn, causing the red vehicle to brake again to prevent a rear-end collision, with a minimum distance of $7$ meters ($md_{t3}=7$). Finally, at time point 4, all three vehicles successfully navigate through the $T$-junction without incident.

Both DriveFuzz and Doppel incorporate the minimum distance ($md$) between vehicles as a key metric in their fitness functions. For example, DriveFuzz evaluates fitness based on the minimum distance and the number of hard braking ($hb$) events by the ego vehicle, while Doppel considers the minimum distance between all vehicles and integrates additional metrics using the NSGA-II algorithm~\cite{zitzler1999multiobjective}. However, both methods rely on aggregate functions, such as \emph{minimum} or \emph{count}, to evaluate the overall scenario fitness. As a result, DriveFuzz computes the fitness for the red vehicle as $hb=2$ and $md=\min(md_{t2}, md_{t3})=5$, while Doppel similarly incorporates $md=5$. These approaches overlook the potential collision risk at time point 3, where the red and green vehicles maintain a distance of $7$ meters ($md_{t3}=7$).

Furthermore, DriveFuzz only counts the number of hard braking events, without considering the spatial or temporal context of these events. By reducing such events to simple numerical values, it fails to capture critical details that could indicate higher-risk scenarios.

In summary, the complex scenario illustrated in Figure~\ref{fig:me} should be prioritized for mutation to further explore potential risks. However, due to the reliance on insufficient aggregate functions (\cf the first limitation), existing methods may assign the same or even lower fitness scores to such scenarios compared to simpler ones, leading to missed opportunities in identifying safety-critical behaviors of ADS.

Furthermore, dangerous or violating behaviors usually occur during interactions with other traffic participants. With more frequent interactions comes a higher risk of violations~\cite{9827305}. Consider another scenario shown in Figure~\ref{fig:me}(b), where the white vehicle is driving away from the other two vehicles. As the interactions decrease, it becomes less likely to exhibit further violating behaviors. Consequently, the white vehicle should be prioritized for mutation to increase interaction. However, both DriveFuzz and Doppel randomly select vehicles for mutation (\cf the second limitation), which inevitably results in redundant offspring scenarios.

To address these challenges, we propose \tool, a fuzz testing method that employs a Transformer encoder to effectively analyze the temporal dynamics within scenarios and evaluate the overall scenario using violation prediction models. Additionally, \tool integrates distance-guided mutation strategies to enhance the likelihood of interactions between vehicles, thereby improving the quality and relevance of offspring scenarios.
\section{Approach}
\label{sect:approach}

\begin{figure}[h]
	\centering
	\includegraphics[width=0.9\linewidth]{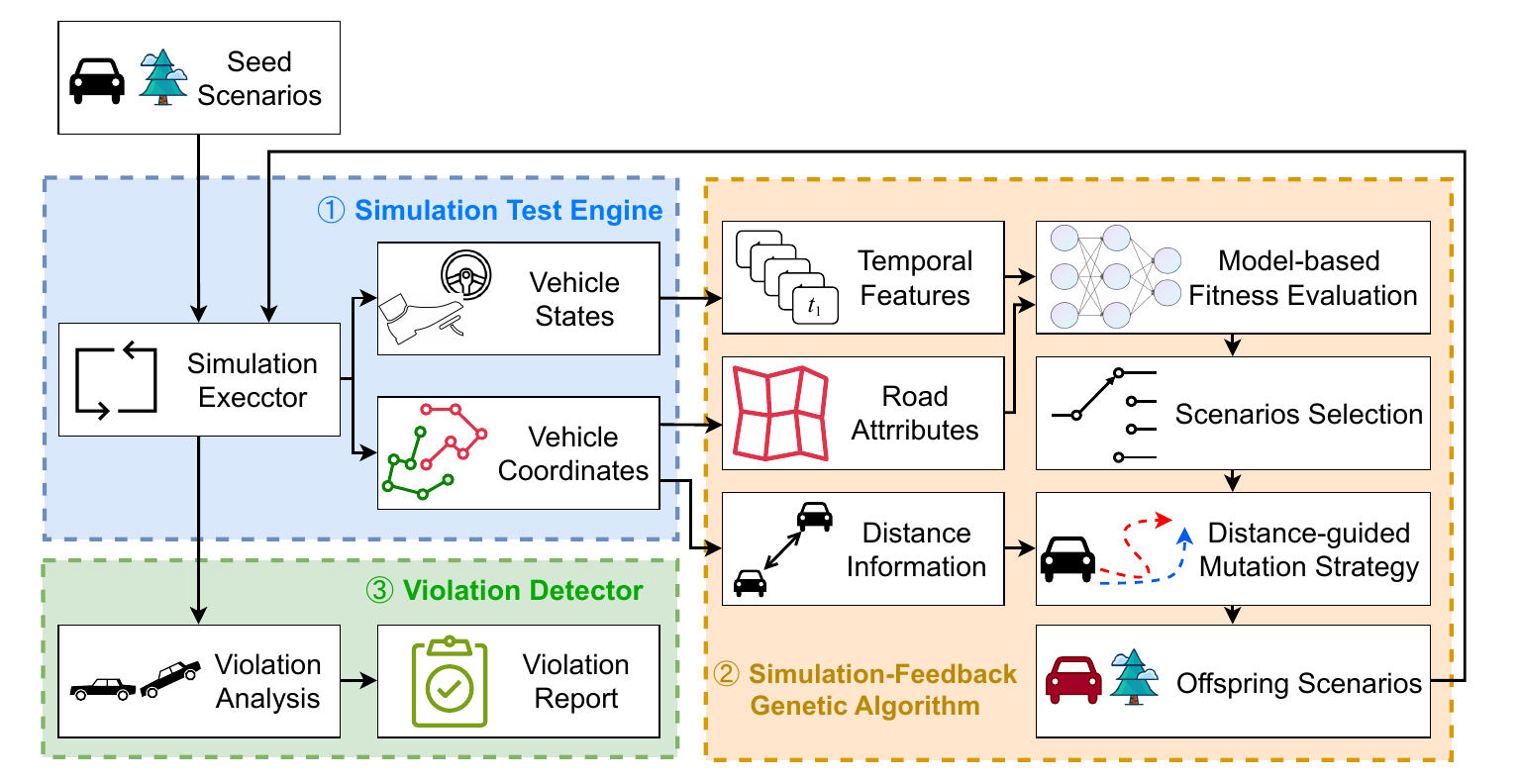}
	\caption{Overview of SimADFuzz}
	\label{fig:simadfuzz}
\end{figure}

Figure~\ref{fig:simadfuzz} presents an overview of \tool, which comprises three modules: the simulation test engine, the simulation-feedback genetic algorithm, and the violation detector.

First, \tool sends seed scenarios to the simulation test engine for execution. The simulation engine collects feedback information, including the coordinates and physical states of vehicles, as detailed in \S~\ref{sec:simulation_test_engine}.

Second, the simulation-feedback genetic algorithm processes the collected feedback information to extract temporal features and road attributes. It then evaluates the fitness score based on the violation prediction model and SDC-Scissor (described in \S~\ref{sec:model-based}). The fitness scores guide the selection of scenarios, which are subjected to crossover (\S~\ref{sec:crossover_strategy}) and mutation (\S~\ref{sect:mutation_strategy}) to generate offspring scenarios that promote more interactions between vehicles and increase the likelihood of triggering violation behaviors.

Finally, the violation detector analyzes the interactions between vehicles, pedestrians, and traffic lights to identify five types of violations. Violation reports, along with the corresponding scenarios that can reproduce the violations, are output to assist ADS developers and testers in identifying defects, as detailed in \S~\ref{sec:violation_detector}.

In the following section, we present the key components of \tool in detail.

\subsection{Simulation Test Engine and Feedback Collection}
\label{sec:simulation_test_engine}

To implement simulation-based testing, we design a simulation test engine that incorporates an autonomous vehicle simulator and a simulation-feedback collector.

The autonomous vehicle simulator is maintained by high-fidelity platforms (\eg CARLA~\cite{dosovitskiy2017carla}, LGSVL~\cite{rong2020lgsvl}), or simulation tools designed for specific ADS (\eg Dreamview~\cite{Dreamview}, AWSIM~\cite{AWSIM}). These simulators provide realistic environments, deploy vehicles and pedestrians at specific coordinates, and return sensor data such as RGB cameras, radar, GPS, and inertial measurement units.

Among these, CARLA and LGSVL have emerged as two widely used platforms in fuzz testing methods. However, LGSVL ceased maintenance and updates in 2022, limiting access to certain maps and assets. In contrast, CARLA has continuously improved its maps and API integrations, offering more comprehensive simulation capabilities that are well-suited for fuzz testing. Additionally, CARLA hosts official ADS performance benchmarks through the CARLA Leaderboard, making it a preferred choice for both research and industry. Consequently, \tool prioritizes CARLA as the primary simulation environment for generating driving scenarios, deploying ADS, and conducting fuzz testing.

The simulation-feedback collector gathers real-time feedback information by calling APIs provided by the simulator. Specifically, \tool collects two types of vehicle-related information, as shown in Table~\ref{tab:simadfuzz-infos}, \ie (1) vehicle coordinates, which indicate the vehicle's location and trajectory; and
(2) physical states, which reflect the vehicle's behavior, including attributes such as speed and acceleration.
This information captures the temporal features of vehicles during simulation and is used to evaluate scenario fitness through a model-based approach.

\begin{table}[h]
	\caption{Feedback Information collected by \tool}
	\centering
	\begin{tabular}{lll}
		\toprule
		\textbf{Type}                        & \textbf{Name} & \textbf{Description}                        \\
		\hline
		\multirow{2}{*}{Vehicle coordinates} & $loc_x$       & X-coordinate of the ego vehicle             \\
		                                     & $loc_y$       & Y-coordinate of the ego vehicle             \\ \hline
		\multirow{5}{*}{Physical states}      & $speed_x$     & X-direction speed of the ego vehicle        \\
		                                     & $speed_y$     & Y-direction speed of the ego vehicle        \\
		                                     & $acc_x$       & X-direction acceleration of the ego vehicle \\
		                                     & $acc_y$       & Y-direction acceleration of the ego vehicle \\
		                                     & $yaw$         & Heading of the ego vehicle                  \\
		\bottomrule
	\end{tabular}
	\label{tab:simadfuzz-infos}
\end{table}

\subsection{Simulation-Feedback Genetic Algorithm}

\begin{figure}[h]
	\centering
	\includegraphics[width=1\linewidth]{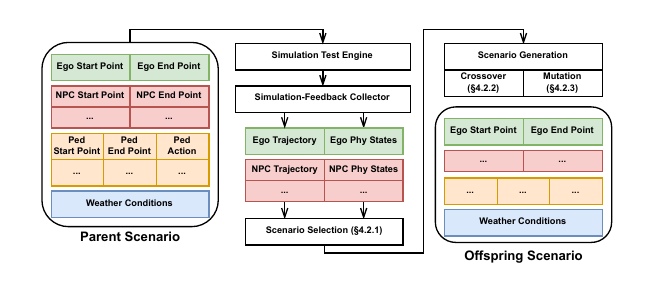}
	\caption{Process of the Simulation-Feedback Genetic Algorithm}
	\label{fig:indv}
\end{figure}

Figure~\ref{fig:indv} illustrates the process of the simulation-feedback genetic algorithm. In \tool, the genetic representation (\ie chromosome) of a scenario used for simulation is composed of four parts. The green and red part shown in Figure~\ref{fig:indv} represent the routes of the ego vehicle and NPC vehicles, respectively, defined by their start and end points. The yellow part represents the routes of pedestrians, while the blue part encodes weather conditions, such as rain or fog levels.

After the parent scenario completes simulation, \tool collects the coordinates and physical states of vehicles. These collected features are then used in a sequential process involving scenario selection, crossover, and mutation to generate offspring scenarios for the next generation of simulations.

We detail the key components of the simulation-feedback genetic algorithm, including model-based fitness evaluation, scenario selection, crossover strategies, and mutation strategies, in the following sections.

\subsubsection{Model-based Fitness Evaluation and Scenario Selection}
\label{sec:model-based}
In general, a driving scenario is a sequence composed of several scenes, where each scene represents a snapshot of the simulation world~\cite{7313256}. Scenes may include actions, events, and other objects that characterize the driving environment. Based on this definition, we formalize a driving scenario $S$ as a sequence $(s_1, \cdots, s_T)$ of $T$ scenes with fixed time intervals. Each scene $s_t$ is represented as $(v^{t}_1, \cdots, v^t_m)$, where $m$ is the number of vehicles in $S$, and $v^t_i$ refers to the features of vehicle $v_i$, including its coordinates and physical states at scene $s_t$.

However, these features only represent individual scenes within the scenario and fail to capture interactions between vehicles or behaviors such as turning or sudden braking. These interactions are critical for evaluating the potential of a scenario to trigger violations.

To address this limitation, we construct a neural network model based on the Transformer encoder to embed driving scenarios into high-dimensional feature representations. The model extracts temporal relationships and interactions among vehicles through its encoding layers. Based on these feature representations, we design a violation prediction model (VPM) to evaluate the fitness of driving scenarios by predicting the probability of triggering violations. This fitness score guides the selection of scenarios for further optimization and mutation. The VPM consists of two main layers, \ie driving scenario encoder and violation prediction layer.

\textbf{Driving Scenario Encoder.} The driving scenario encoder is designed to learn patterns from scene sequences using sequence models, which are effective in capturing relevant information from temporal data. Sequence models such as LSTM~\cite{hochreiter1997long} and Transformer have been widely applied to tasks like driving behavior intention recognition~\cite{gao2023probabilistic} and trajectory/velocity prediction~\cite{gao2021trajectory, ma2015long, geng2023physics}. In this work, we adopt the Transformer for the driving scenario encoding task due to its superior performance in trajectory forecasting~\cite{9412190, xu2023leveraging}. In \tool, the Transformer encoder takes feedback information as input, represented as a tensor with the shape $T \times N_{info}$. Here, $T$ denotes the number of scenes, and $N_{info}$ represents the total amount of information collected for all vehicles in the scenario. Specifically, $N_{info} = 7 \times m$, where $m$ is the number of vehicles, and the 7 features include coordinates, speed, acceleration, and heading. By leveraging the multi-head attention mechanism, the Transformer encoder embeds the scenario into a high-dimensional feature space, which is then used to predict the probability of violations.

\textbf{Violation Prediction Layer.} The violation prediction layer is designed to estimate the probability of violations, which serves as the fitness score for the scenario. It consists of a fully connected (FC) layer with a single output node, producing a scalar probability value. The FC layer combines and models relationships among features extracted from the entire scene sequence, capturing high-level interactions and temporal dependencies. Scenarios with higher predicted probabilities indicate a greater likelihood of causing violations in the ADS. Consequently, these scenarios are prioritized for selection in generating offspring scenarios during the optimization process.

The VPM evaluates scenarios and prioritizes them for crossover and mutation in the genetic algorithm. However, relying solely on a single fitness metric can be misleading, as it may fail to capture other critical characteristics of the scenarios being tested~\cite{ml2st}. As shown in Table~\ref{tab:fitness}, existing works commonly evaluate scenarios using multiple fitness metrics to ensure a more comprehensive assessment. To address this limitation, we augment the \tool with three additional fitness metrics: minimum distance, number of unique violations~\cite{cheng2023behavexplor,Doppel}, and SDC-Scissor (Self Driving Cars Cost Effective Test Selector)~\cite{Birchler2023}.

\begin{itemize}
	\item The minimum distance is the smallest observed distance between the ego vehicle and other vehicles throughout the scenario simulation. A smaller distance typically indicates a higher likelihood of collisions or other violations, reflecting an increased level of risk in the scenario.
	\item The number of unique violations refers to the total number of distinct violations triggered by the ego vehicle after filtering out duplicates and highly similar violations. A higher count of unique violations suggests a more risky and complex scenario, potentially exposing more defects in the ADS by challenging its performance in diverse ways.
	\item SDC-Scissor is a method that leverages machine learning models to identify and filter unlike scenarios  to detect faults in ADS before executing them. It extracts static road attributes (as shown in Table~\ref{tab:road-attributes}) and sends them into trained models to classify scenarios as safe or unsafe. Complementarily, \tool focuses on capturing temporal features, such as dynamic interactions between vehicles over time. Combining these static and temporal features enhances the overall fitness evaluation by leveraging the strength of both approaches.
\end{itemize}

\begin{table}[h]
	\caption{Name and Description of SDC-Features~\cite{Birchler2023}}
	\centering
	\begin{tabularx}{\textwidth}{lX}
		\toprule
		\textbf{Name}                      & \textbf{Description}                                               \\
		\hline
		Direct distance                    & Euclidean distance between start and finish                        \\
		Length                             & Total length of the driving path                                   \\
		Number L Turns                     & Number of left turns on the driving path                           \\
		Number R Turns                     & Number of right turns on the driving path                          \\
		Number Straight                    & Number of straight segments on the driving path                    \\
		Total angle                        & Cumulative turn angle on the driving path                          \\
		Median, Std, Max, Min, Mean angle  & Median/Std/Maximum/Minimum/Average turn angle on the driving path  \\
		Median, Std, Max, Min, Mean radius & Median/Std/Maximum/Minimum/Average turn radius of the driving path \\
		Full road diversity                & The cumulative diversity of the full road composed of all segments \\
		Mean road diversity                & The mean diversity of the segments of a road                       \\
		\bottomrule
	\end{tabularx}
	\label{tab:road-attributes}
\end{table}

To this end, we combine the VPM with three additional fitness metrics, treating scenario selection as a multi-objective optimization task. The goal is to identify solutions (\ie scenarios) that balance multiple objectives, such as increasing violation probabilities and risk exposure, or reducing the distance between vehicles. To achieve this, we leverage the Non-dominated Sorting Genetic Algorithm II (NSGA-2)~\cite{zitzler1999multiobjective}, a widely used multi-objective optimization algorithm, to select scenarios based on the Pareto-optimal frontier.

NSGA-2 identifies a set of solutions that are not dominated by any other solutions, meaning that no other solutions perform better across all objectives. By employing non-dominated sorting and crowding distance techniques, NSGA-2 ensures a diverse set of Pareto-optimal scenarios, effectively exploring the simulation search space while avoiding premature convergence to local optima.

\subsubsection{Crossover Strategy}
\label{sec:crossover_strategy}

The crossover operator combines two parent scenarios to generate two offspring scenarios. \tool performs the crossover operation by swapping the routes of NPC vehicles and pedestrians, while keeping the other two parts (the route of the ego vehicle and the weather conditions) unchanged.

For the pedestrian swap operation, \tool randomly swaps half of the pedestrians between the two scenarios, increasing the diversity of pedestrian routes and behaviors.

For the NPC vehicle swap operation, \tool is designed to enhance the interaction likelihood between the ego vehicle and NPC vehicles. Specifically, for two parent scenarios $S_1$ and $S_2$, \tool checks whether the trajectory of an NPC vehicle $NPC^{S1}_i$ in $S_1$ intersects with the trajectory of the ego vehicle $EGO^{S2}$ in $S_2$. If it is the case, there might be a potential interaction between $EGO^{S2}$ and $NPC^{S1}_i$, and \tool randomly selects one NPC vehicle from $S_2$ and swaps it with $NPC^{S1}_i$, generating two offspring scenarios.

\subsubsection{Mutation Strategy}
\label{sect:mutation_strategy}

Mutation operators generate offspring scenarios based on a parent scenario. As shown in Figure~\ref{fig:indv}, the chromosome of a scenario is categorized into three components: vehicles (including the ego and NPC vehicles), pedestrians, and weather. Among these, vehicles play a critical role in contributing to dynamic scenario variations. To enhance the likelihood of interactions with the ego vehicle, we focus on vehicle mutation and propose a distance-guided mutation strategy.

\textbf{Distance-guided Mutation for Vehicles.} Vehicles are essential traffic participants in creating dynamic scenarios for testing. As discussed in Section~\ref{sect:example}, \textit{the probability of interactions and subsequent violations decreases as the distance between vehicles increases}. To address this, \tool employs a distance-guided mutation strategy to identify and remove vehicles with low interaction potential. Specifically, two types of vehicles are removed:
\begin{itemize}
	\item Vehicles that remain nearly stuck throughout the simulation, contributing minimally to scenario dynamics.
	\item Vehicles that consistently move away from other vehicles, reducing the likelihood of interactions.
\end{itemize}

The process of distance-guided mutation is detailed in Algorithm~\ref{algo:mutate}. The algorithm takes the following inputs:
the set of NPC vehicles in the parent scenario ($C$), the ego vehicle in the parent scenario ($e$), the distance matrix ($m_{dis}$), a threshold for identifying stuck vehicles ($w$), and a time window size for identifying leaving vehicles ($u$). The algorithm outputs a modified set of NPC vehicles for the offspring scenario.

The $m_{dis}$ is represented as a three-dimensional matrix with shape $T \times |V| \times |V|$, where $T$ denotes the number of time steps, and $|V|$ represents the number of vehicles. Each element \(m_{dis}(t, v_1, v_2)\) represents the Euclidean distance between vehicles \(v_1\) and \(v_2\) at timestamp \(t\). The distance is computed as:
\begin{equation*}
	m_{dis}(t, v_1, v_2) = \begin{cases}
		\|v_1^t - v_2^t\|_2, & \text{if } v_1 \neq v_2 \\
		0,                   & \text{otherwise}
	\end{cases}
\end{equation*}
where \(\|v_1^t - v_2^t\|_2\) denotes the Euclidean distance between the positions of vehicles \(v_1\) and \(v_2\) at time \(t\).

\begin{algorithm}[h]
	\caption{Distance-guided Mutation Process}
	\label{algo:mutate}
	\SetKwInOut{Input}{Input}
	\SetKwInOut{Output}{Output}

	\Input{
		$C$ - NPC vehicles in the parent scenario, \\
		$e$ - Ego vehicle in the parent scenario, \\
		$m_{dis}$ - Distance matrix, \\
		$w$ - Threshold for identifying stuck vehicles, \\
		$u$ - Time window size for identifying leaving vehicles
	}
	\Output{
		$C$ - Mutated NPC vehicles
	}

	$D \gets \emptyset$\;

	\For{$v \in C$}{
	$\Delta_{route} \gets \sum_{t=1}^{T} \|v^t - v^{t-1}\|_2$ \;  
	\If{$\Delta_{route} < w$}{
		$D \gets D \cup \{v\}$\; 
		\textbf{continue}\; 
	}

	$flag \gets True$ \;
	\For{$t \in [u, T]$}{
	$\Delta_{dis} \gets \sum_{i=t-u}^{t} (m_{dis}(i+1, v, e) - m_{dis}(i, v, e))$\;
	\If{$\Delta_{dis} < 0$}{
		$flag \gets False$\;
		\textbf{break}\;
	}
	}
	\If{$flag$}{
		$D \gets D \cup \{v\}$\;
	}
	}
	$C \gets C \setminus D$ \;
	\textsc{Mutate\_Route}($C$) \;
	\textsc{Mutate\_Route}($e$) \;
	\textsc{Add\_Vehicles}($C$, $|D|$) \;
	\Return{$C$}\;
\end{algorithm}

Algorithm~\ref{algo:mutate} begins by initializing an empty set \(D\) to record NPC vehicles marked for removal. Each vehicle \(v \in C\) is evaluated based on two criteria:

1. \textbf{Stuck Vehicles}: The algorithm computes the total route length \(\Delta_{route}\) for each vehicle \(v\) by summing the Euclidean distances between its positions at consecutive timestamps. If \(\Delta_{route} < w\), indicating minimal movement throughout the simulation, the vehicle is added to \(D\).

2. \textbf{Leaving Vehicles}: For each vehicle \(v\), the algorithm computes the cumulative distance trend \(\Delta_{dis}\) relative to the ego vehicle \(e\) over a sliding time window of size \(u\). If \(\Delta_{dis}\) remains positive (\(\geq 0\)) across all examined windows, the vehicle is marked as consistently moving away and added to \(D\).

Vehicles in \(D\) are removed from the scenario. The function \textsc{Mutate\_Route} modifies the routes of the remaining vehicles and the ego vehicle \(e\), ensuring the mutated routes remain on the same road segment for contextual consistency. The function \textsc{Add\_Vehicles} then replenishes the removed vehicles with new NPC vehicles, generating routes with randomly selected start and end points to explore new routes. The final mutated set of NPC vehicles is returned as part of the offspring scenario.

\begin{figure}[h]
	\centering
	\includegraphics[width=1\linewidth]{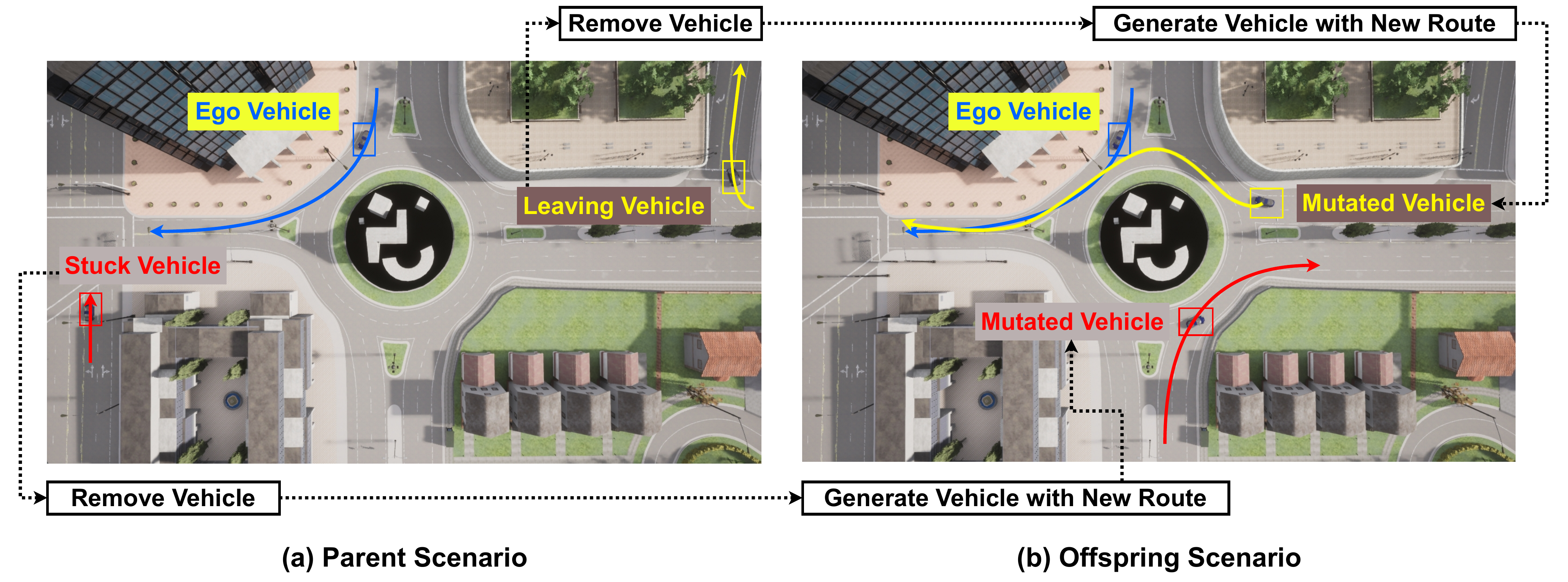}
	\caption{An Example of Distance-guided Mutation}
	\label{fig:mutate_example}
\end{figure}

Figure~\ref{fig:mutate_example} illustrates an example of distance-guided mutation. In the parent scenario (left), the red vehicle is identified as a \textit{stuck vehicle}, as it remains stationary at a long red traffic light, while the yellow one is identified as a \textit{leaving vehicle}, as its trajectory indicates it is consistently moving away from the ego vehicle (the blue vehicle), reducing the likelihood of interactions. These two vehicles are removed during the mutation process.

In the offspring scenario (right), two new vehicles are added to replace the removed ones. One of the new vehicles intersects with the ego vehicle's trajectory at a roundabout, thereby enhancing the potential for interactions in the offspring scenario.

\textbf{Mutation for Pedestrians and Weather Conditions.} \tool also mutates pedestrians and weather conditions. Pedestrians are spawned near the ego vehicle to improve interaction likelihood, as their relatively low speed makes distant interactions unlikely. Weather conditions are randomly sampled from predefined ranges, such as rain intensity (0--100) and sun altitude (-90° to 90°), to explore diverse environmental scenarios.

\subsection{Violation Detector and Reproduction}
\label{sec:violation_detector}

The violation detector module is responsible for identifying misbehavior and violations triggered by ego vehicles during simulation. \tool supports detecting the following violations:

\textbf{Collision:} Collisions are one of the most fundamental violations that an ADS must avoid. Collisions are detected when the ego vehicle comes into physical contact with other static or dynamic objects. This detector is implemented using CARLA's built-in collision sensors.

\textbf{Lane Invasion:} Lane invasions are detected when the ego vehicle crosses restricted road lanes, such as crossing solid lines that indicate non-crossable lane boundaries during lane changes. This detector is implemented using CARLA's built-in lane invasion sensors.

\textbf{Speeding:} Speeding violations are detected when the ego vehicle's speed consistently exceeds the specific road's speed limit for a duration $T_{speeding}$. This detector is implemented by analyzing the ego vehicle's historical speed data and comparing it against the speed limit.

\textbf{Running Red Lights:} Running red light violations are detected when the ego vehicle drives through an intersection during a red light. This detector is implemented by monitoring the traffic light state and the ego vehicle's coordinates within the intersection.

\textbf{Stuck:} Stuck behaviors often indicate ADS failures or system disablement but may also result from external factors such as traffic congestion. These violations are detected when the ego vehicle remains stationary beyond a predefined duration $T_{stuck}$. This detector is implemented by analyzing the ego vehicle's historical speed data and verifying whether the speed remains consistently zero over the duration $T_{stuck}$.

When \tool identifies a violation, it saves the entire scenario (including the states of the ego vehicle, NPC vehicles, pedestrians and weather conditions) to facilitate reproduction. Additionally, \tool utilizes CARLA's API (\texttt{Client.start\_recorder}) to record scenarios with more detailed information (such as the states of traffic lights and vehicle dynamics). \tool supports replaying both types of scenario recordings to reproduce violations, and supports further analysis, including root cause investigation and ADS performance evaluation.
\section{Evaluation and Results}
\label{sect:exp}

To evaluate the effectiveness of \tool, we conducted experiments aimed at addressing the following research questions, covering various perspectives:

\begin{description}
	\item[RQ1] To what extent can optimization strategies in scenario selection and mutation components of \tool improve its effectiveness in detecting violations?
	\item[RQ2] How effective is \tool in detecting violations in ADS compared to state-of-the-art fuzzers?
	\item[RQ3] Does the strategy used in \tool impact the diverse generation of scenarios?
\end{description}

\subsection{Experimental Settings}
\subsubsection{ADS Under Test}
The ADS system evaluated in our experiments is InterFuser~\cite{shao2023safety}. InterFuser utilizes a Transformer-based architecture for interpretable sensor fusion, integrating sensor data to generate control commands for the ego vehicle. It has demonstrated its effectiveness in autonomous driving tasks by achieving high driving scores on the CARLA Leaderboard. Additionally, InterFuser is open-source, and the weights of its models are publicly available. These features make it a suitable choice for evaluating the performance of \tool.

\subsubsection{Fuzzing Configurations}
To evaluate the scenarios generated by \tool, we utilize the CARLA simulator, configured to run at a frame rate of 20 Hz. Each simulation lasts up to 10 minutes but may terminate earlier if the ego vehicle collides with other vehicles or pedestrians, or if it successfully reaches its destination.

The simulation environment is based on the Town03 map provided by CARLA, which represents a large urban area resembling a downtown district. This map includes roundabouts, underpasses, overpasses, and other complex road structures, offering diverse and challenging driving scenarios for testing.

\begin{figure}[h]
	\centering
	\begin{subfigure}{0.32\textwidth}
		\centering
		\includegraphics[height=3cm]{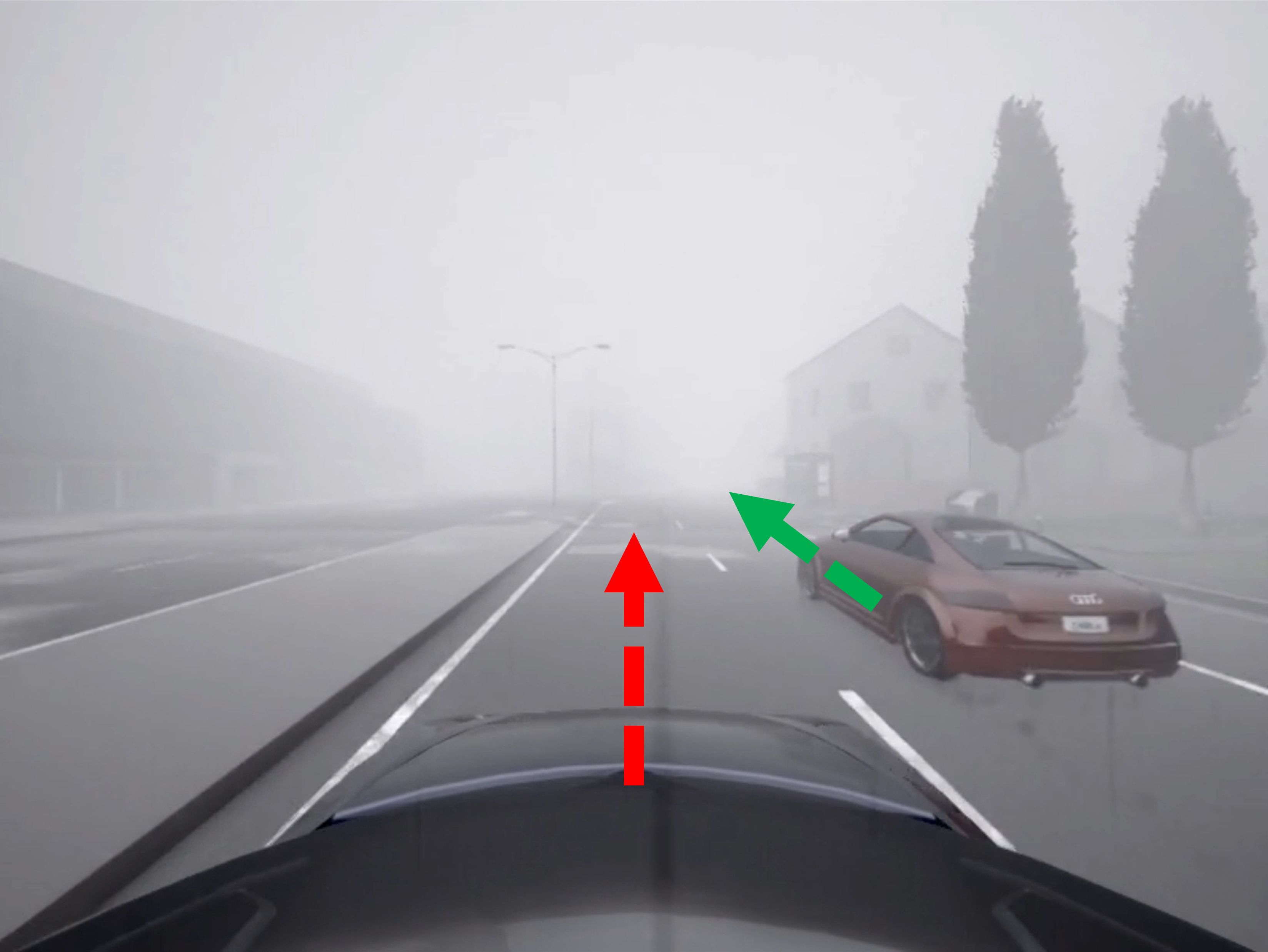}
		\caption{Two vehicles driving in the same direction}
	\end{subfigure}
	\hfill
	\begin{subfigure}{0.32\textwidth}
		\centering
		\includegraphics[height=3cm]{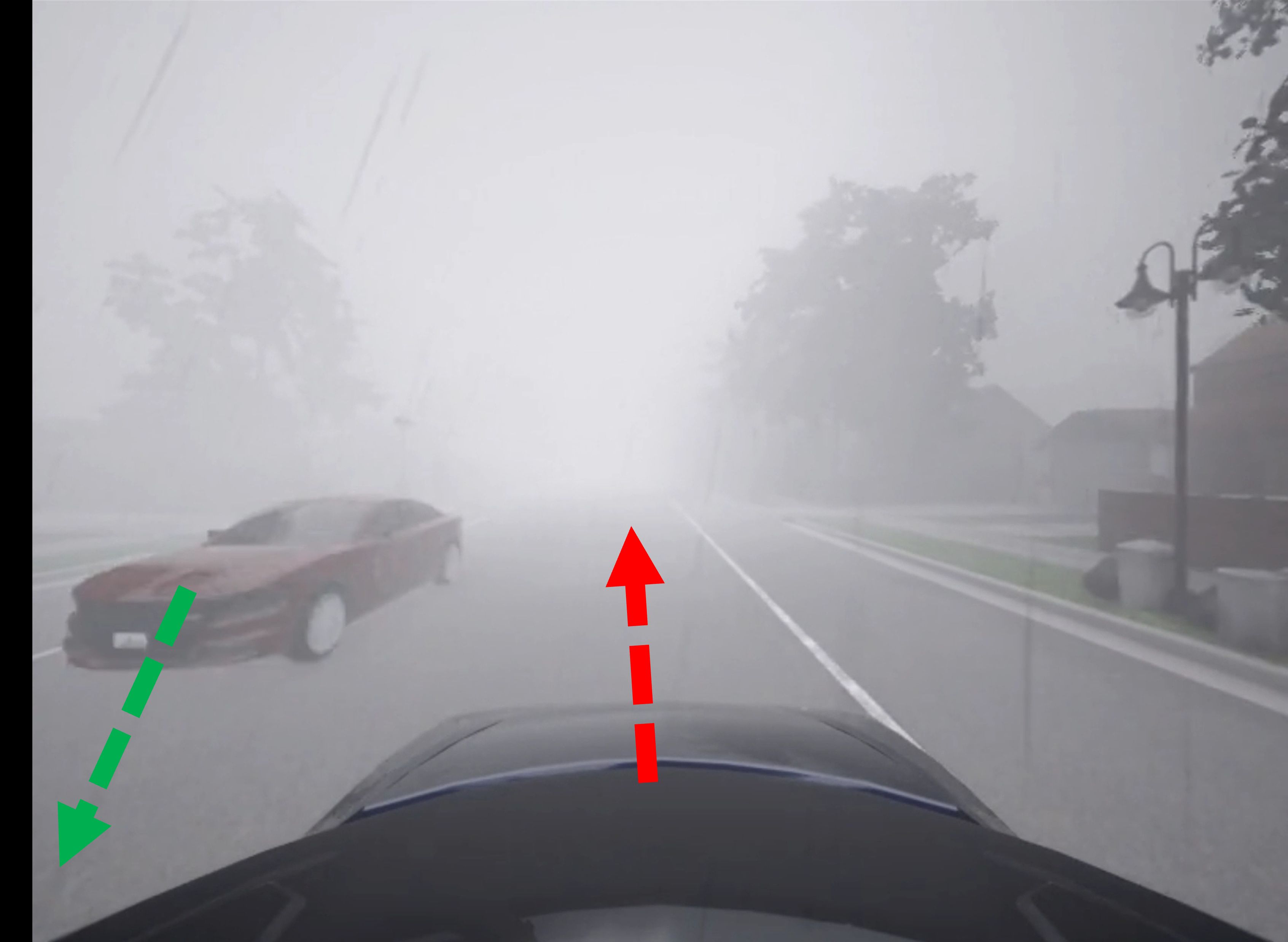}
		\caption{Two vehicles driving in the opposite direction}
	\end{subfigure}
	\hfill
	\begin{subfigure}{0.32\textwidth}
		\centering
		\includegraphics[height=3cm]{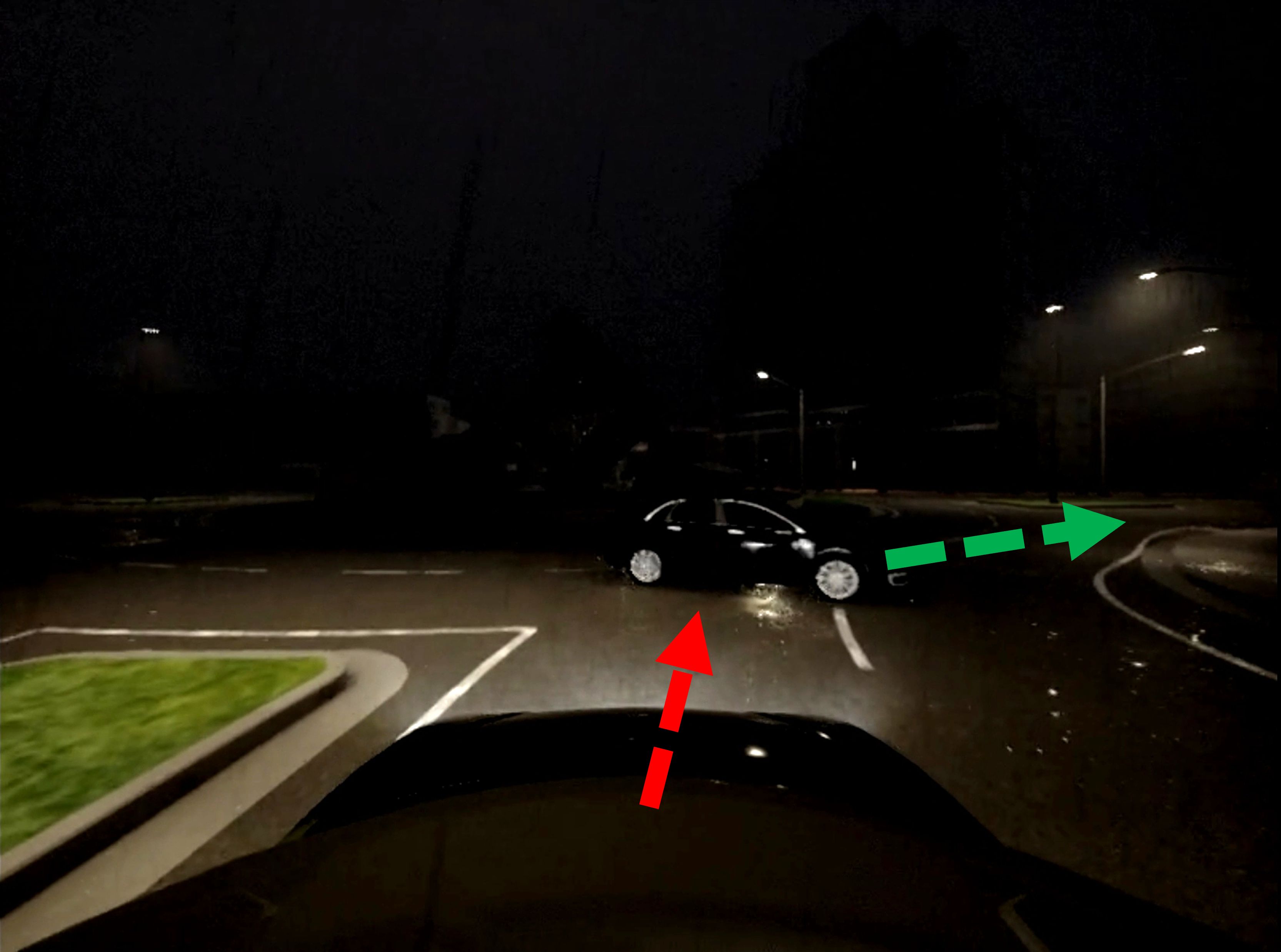}
		\caption{Encountering vehicles at the roundabout}
	\end{subfigure}
	\vspace{0.2cm}
	\begin{subfigure}{0.32\textwidth}
		\centering
		\includegraphics[height=3cm]{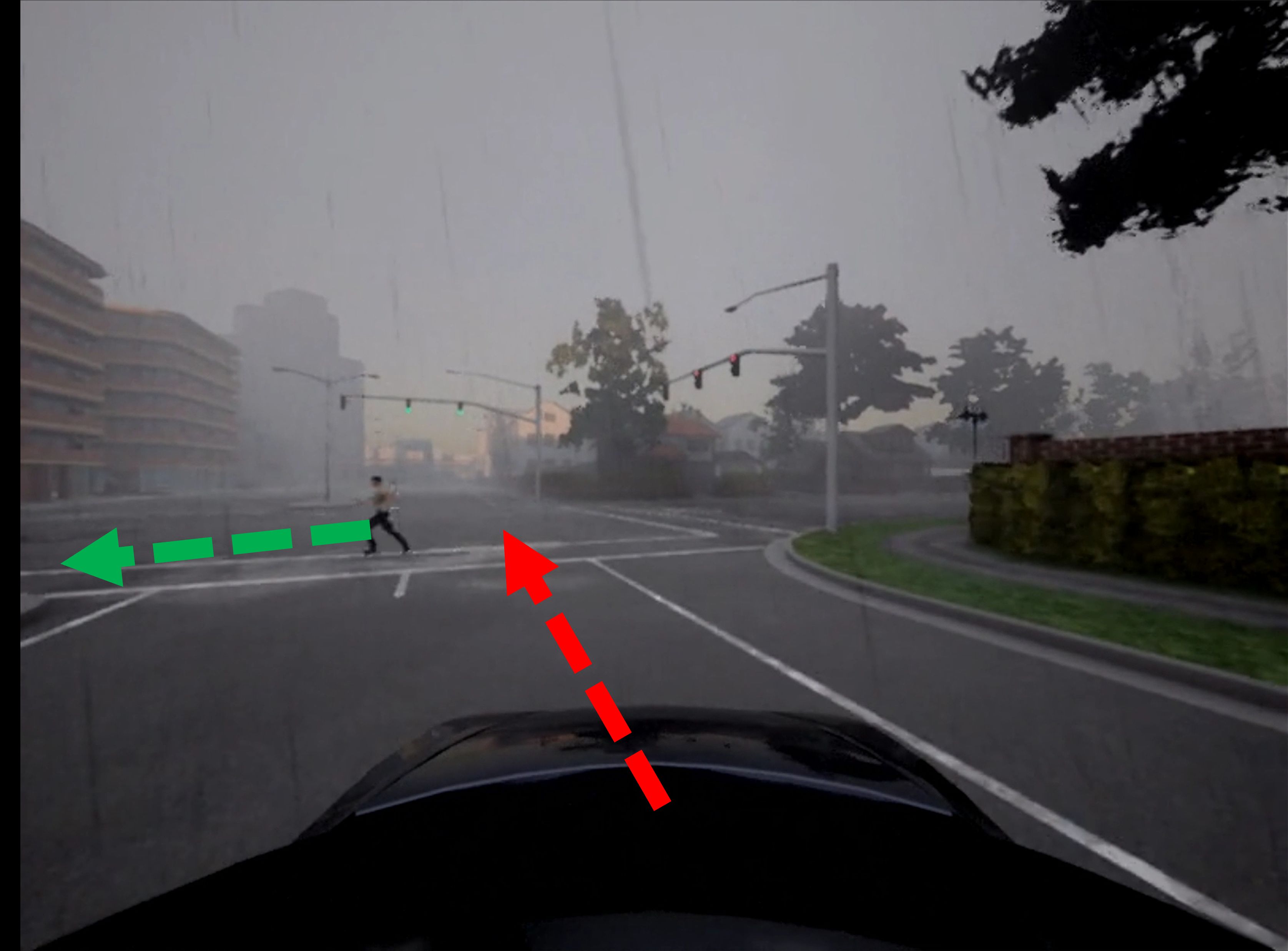}
		\caption{Pedestrians crossing intersections}
	\end{subfigure}
	\hfill
	\begin{subfigure}{0.32\textwidth}
		\centering
		\includegraphics[height=3cm]{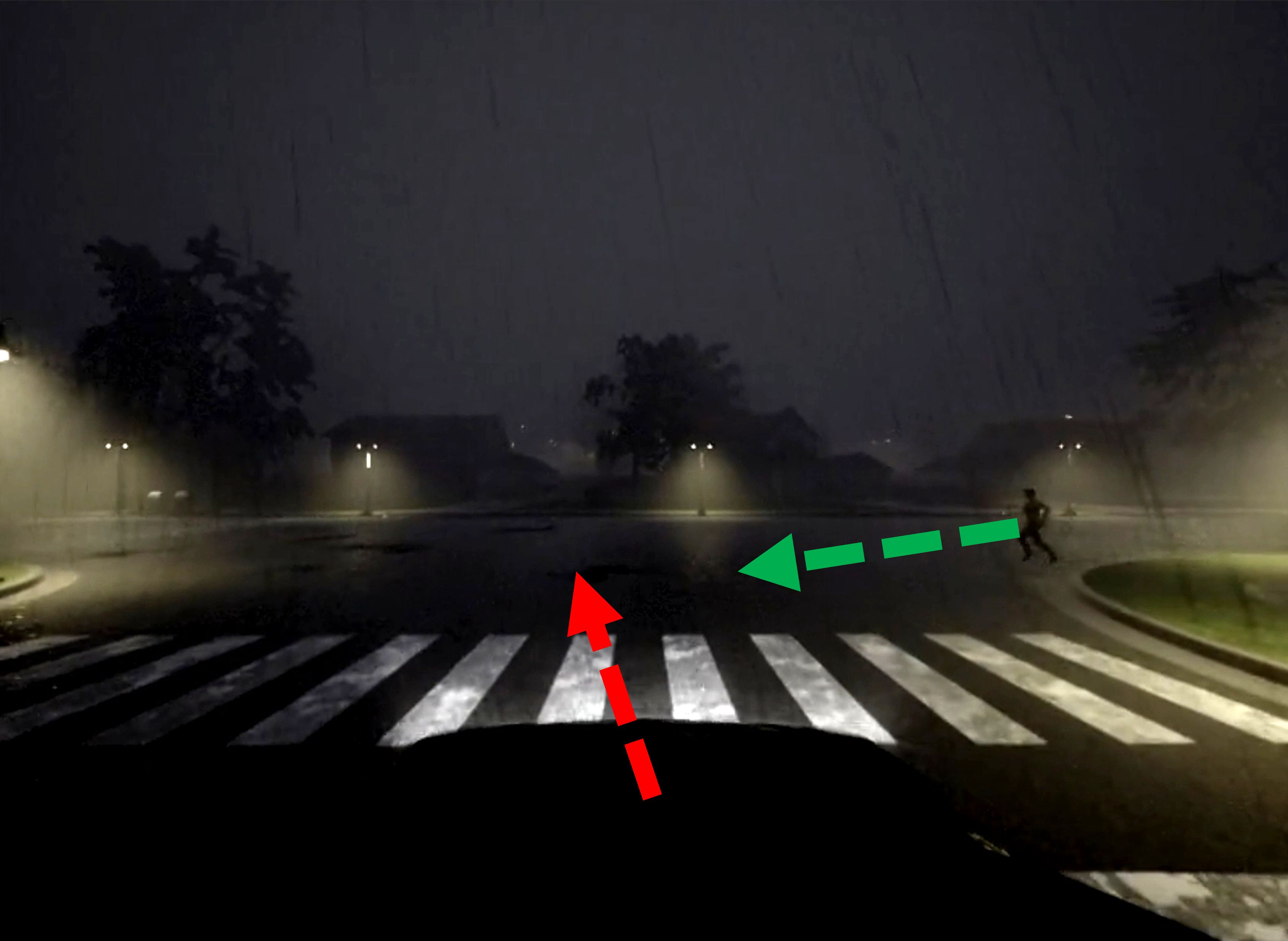}
		\caption{Pedestrians crossing intersections in low-light conditions}
	\end{subfigure}
	\hfill
	\begin{subfigure}{0.32\textwidth}
		\centering
		\includegraphics[height=3cm]{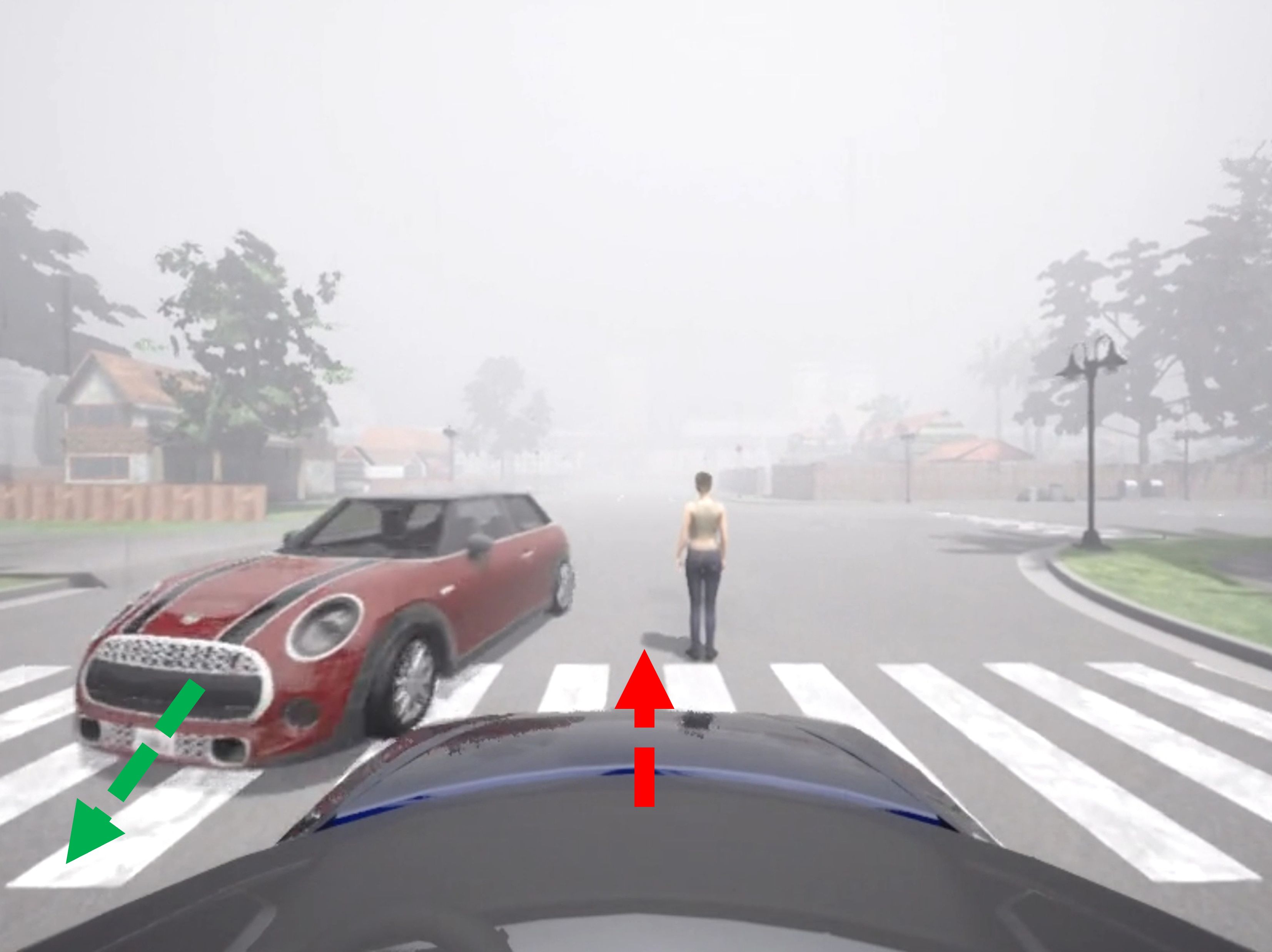}
		\caption{Multiple participants meeting at an intersection}
	\end{subfigure}
	\caption{Scenarios Generated During the Fuzz Testing}
	\label{fig:scenarios}
\end{figure}

Each testing scenario includes ten pedestrians and three vehicles, one of which is configured as the ego vehicle deployed with InterFuser. Figure~\ref{fig:scenarios} illustrates six typical scenarios generated by \tool: two vehicles driving in the same or opposite directions, vehicles encountering each other at the roundabout, pedestrians crossing intersections under different lighting conditions, and multiple participants meeting at an intersection.

The population size for fuzzing is set to 20, with crossover and mutation probabilities fixed at 0.5. To ensure a fair comparison across different fuzzing methods, the same seed scenarios are utilized for all experiments. These seed scenarios are generated by randomly selecting ego vehicle routes from the Town03 map.

There are several hyperparameters in \tool, as shown in Table~\ref{tab:hyper}. The first three are related to the violation prediction model (VPM), including parameters for the Transformer encoder. The remaining three are associated with mutation strategies, including two parameters mentioned in Algorithm~\ref{algo:mutate} and the maximum distance between pedestrians and the ego vehicle when generating pedestrians.

\begin{table}[h]
	\caption{Hyperparameters of \tool}
	\label{tab:hyper}
	\resizebox{\textwidth}{!}{
		\begin{NiceTabular}{c|c|c}
			\toprule
			\textbf{Category}                 & \textbf{Hyperparameter Description}                     & \textbf{Value} \\ \midrule
			\multirow{3}{*}{VPM-related}      & embedding dimension                                     & 128            \\
			                                  & head number                                             & 3              \\
			                                  & encoder layer number                                    & 3              \\ \midrule
			\multirow{3}{*}{Mutation-related} & threshold for identifying stuck vehicles                & 10 meters      \\
			                                  & time window size for identifying leaving vehicles       & 10 seconds     \\
			                                  & maximum distance between pedestrian and the ego vehicle & 20 meters      \\ \bottomrule
		\end{NiceTabular}%
	}
\end{table}

We generate 1,000 simple and short driving scenarios, each with a simulation time constrained to 2 minutes, and collect feedback information to train the violation prediction model and the SDC-Scissor model. The SDC-Scissor model is trained using a Logistic Regression classifier, which has been shown to effectively performance in the study by Birchler et al.~\cite{birchler2022cost}. For the violation prediction model, the Adam optimizer is used with a learning rate of 0.001, and binary cross-entropy is employed as the loss function. To mitigate overfitting, an early stopping strategy with a patience threshold of 10 epochs is implemented.

\subsubsection{Baselines and Evaluation Metrics}

We compare \tool against three different fuzzers:

\textbf{AV-Fuzzer}, which employs genetic algorithms to minimize the ego vehicle's safety potential. It generates offspring scenarios by altering the positions of NPCs.

\textbf{DriveFuzz}, which designs a fitness function based on hard acceleration and other behaviors to evaluate scenarios. It mutates the weather conditions, NPC positions, and pedestrian navigation types to generate diverse driving scenarios.

\textbf{TM-Fuzzer}, which dynamically manages traffic flow to increase interactions with the ego vehicle. It also incorporates clustering analysis to generate diverse test scenarios.

We use the number of unique violations (UVs) as the evaluation metric. UVs are defined as violations that occur at different times or locations. This metric is widely used in ADS fuzz testing~\cite{cheng2023behavexplor,Doppel} and serves as a reliable indicator of a fuzzer's performance in detecting violations.

In our evaluation, we define a temporal threshold of ±10 seconds and a spatial threshold of ±30 meters to determine unique violations.

\subsection{Experimental Results}

\subsubsection{RQ1: Components Effective}

To validate the effectiveness of the violation prediction model, SDC-Scissor, and distance-guided mutation strategies, we conducted ablation experiments on several variants of \tool. Each variant was fuzzed for a total of 3 hours.

\begin{figure}[h]
	\centering
	\includegraphics[width=0.9\linewidth]{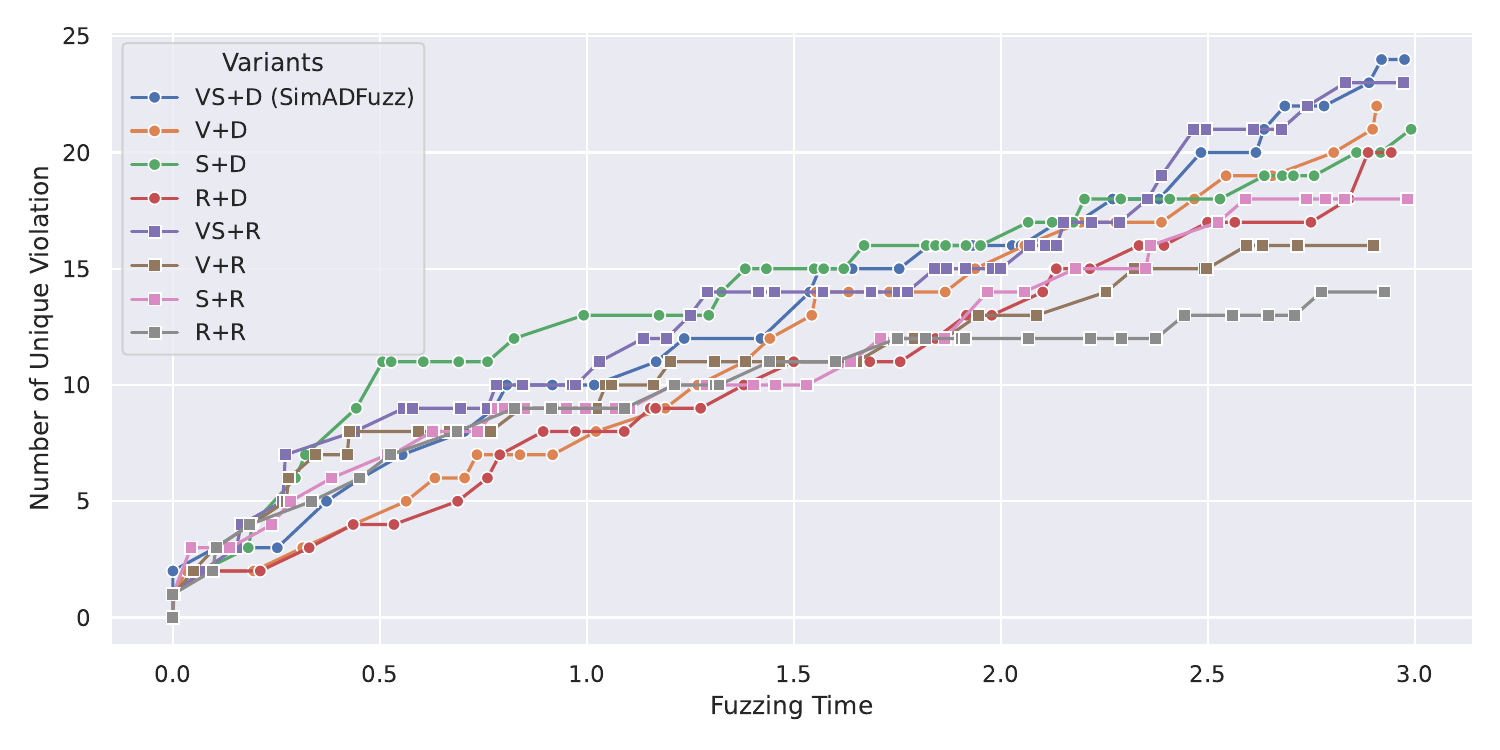}
	\caption{The number of UVs detected by variants of \tool. Each variant is denoted as \( X + Y \), where \( X \) represents the components activated during scenario selection (with \( V \) for the violation prediction model, \( S \) for SDC-Scissor, and \( R \) for random selection), and \( Y \) represents the components activated during mutation (with \( D \) for distance-guided mutation strategies and \( R \) for random mutation strategies).}
	\label{fig:rq1-lane}
\end{figure}

The results are shown in Figure~\ref{fig:rq1-lane}, where each line represents a variant $X+Y$. Here, $X$ indicates the components activated during scenario selection, which can be $VS$ (using both the violation prediction model and SDC-Scissor, representing the complete fuzzer of \tool), $V$ (using only the violation prediction model), $S$ (using only SDC-Scissor), or $R$ (randomly selecting scenarios). $Y$ indicates the components activated during mutation, which can be $D$ (using distance-guided mutation strategies) or $R$ (randomly mutating scenarios).

As shown in Figure~\ref{fig:rq1-lane}, the complete fuzzer $VS+D$ outperforms all other variants, discovering 24 UVs over a total of 3 hours of fuzzing, while the random baseline ($R+R$) only discovers 14 UVs. This indicates that the optimization components for both scenario selection and mutation greatly improve the ability to detect violations in ADS.

Focusing on scenario selection, $V+D$ and $S+D$ outperform $R+D$, indicating that both the violation prediction model and SDC-Scissor improve the effectiveness of selecting high-risk scenarios. The combination of these two models ($VS+D$) further refines the selection process by leveraging temporal features from the violation prediction model and static road features from SDC-Scissor.

In terms of scenario mutation, variants with random mutation strategies ($VS+R$, $V+R$, $S+R$, and $R+R$) perform worse than their counterparts with distance-guided mutation strategies. Furthermore, $VS+R$ shows a stagnation in the number of unique violations detected during fuzzing. This suggests that random mutation strategies may generate low-quality scenarios with limited interaction between traffic participants. Table~\ref{tab:rq1} supports this observation by showing that distance-guided mutation increases the number of NPC vehicles near the ego vehicle (\ie within 50 meters). This increased proximity enhances interactions, influencing the ADS's perception and decision-making, and ultimately leading to more detected violations.

\begin{table}[h]
	\caption{The number of NPC vehicles near by the ego vehicle}
	\label{tab:rq1}
	\begin{NiceTabular}{c|c|c}
		\toprule
		\textbf{Fuzzer variants}          & \textbf{Fuzzing Time} & \textbf{Num of Vehicle} \\ \midrule
		\multirow{3}{*}{VS+R}             & 1h                    & 7                       \\
		                                  & 2h                    & 16                      \\
		                                  & 3h                    & 23                      \\ \midrule
		\multirow{3}{*}{VS+D (SimADFuzz)} & 1h                    & 8                       \\
		                                  & 2h                    & 25                      \\
		                                  & 3h                    & 35                      \\ \bottomrule
	\end{NiceTabular}%
\end{table}

\begin{tcolorbox}[]
	{ \textbf{Answer to RQ1: } The model-based fitness evaluation and distance-guided mutation strategies significantly enhance the performance of \tool, detecting 10 more unique violations compared to variants using random strategies over 3 hours fuzz testing.}
\end{tcolorbox}

\subsubsection{RQ2: Performance Compared to SOTA Fuzzers}

We conducted experiments to compare the performance of \tool with other state-of-the-art fuzzers, including AV-Fuzzer, DriveFuzz, and TM-Fuzzer. Each fuzzer was executed for a total of 6 hours.

\begin{figure}[h]
	\centering
	\includegraphics[width=0.9\linewidth]{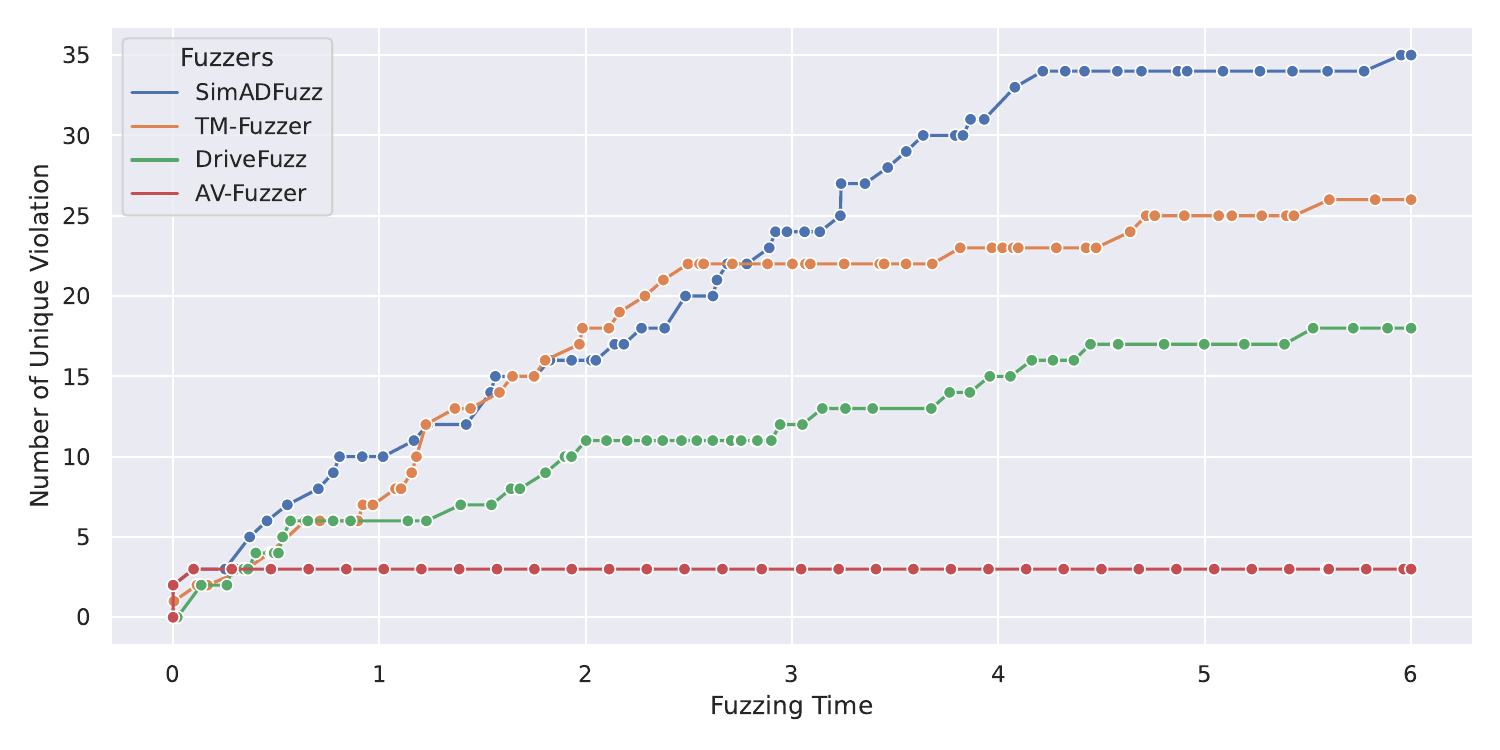}
	\caption{The number of UVs detected by \tool and baselines}
	\label{fig:rq2-line}
\end{figure}

The results are shown in Figure~\ref{fig:rq2-line}, illustrating that the performance gap between \tool and other fuzzers becomes increasingly significant as fuzzing progresses. After 6 hours, \tool detected 35 UVs, outperforming TM-Fuzzer (26 UVs), DriveFuzz (18 UVs), and AV-Fuzzer (3 UVs). Additionally, \tool identified its first collision violation within 17 minutes, compared to 21 minutes for TM-Fuzzer, demonstrates \tool's superior efficiency in detecting critical violations. Specifically, \tool detected 35 UVs, including 4 collisions, 20 lane invasions, and 11 stuck violations.

Notably, none of the fuzzers detected speeding or red light violations during testing with InterFuser. We attribute this to InterFuser's conservative configuration, including a strict speed limit of 5 m/s and a confidence threshold of 0.3 for detecting red traffic lights. These configurations make the ego vehicle behave as a cautious driver, avoiding overspeeding or moving when uncertain about the traffic light state.

To further analyze the results, we categorized the violations discovered by \tool according to their types. Table~\ref{tab:rq2} summarizes six types of violations, briefly describing their scenarios and participants. Importantly, all scenarios were confirmed to be reproducible using \tool.

\begin{table}[h]
	\caption{Violation Types and Scenario Descriptions}
	\label{tab:rq2}
	\resizebox{\textwidth}{!}{%
		\begin{NiceTabular}{c|l|l}
			\toprule
			\textbf{Violation Types}       & \textbf{ID} & \textbf{Scenarios Description}                                                                         \\ \midrule
			\multirow{4}{*}{Collision}     & \#1         & Ego vehicle failed to respond appropriately to a pedestrian jaywalking, resulting in a side collision. \\
			                               & \#2         & Ego vehicle collides with another vehicle while changing lanes to exit a crossroads.                   \\
										   & \#3         & Ego vehicle collides with another vehicle while changing lanes at an intersection.                   \\
			                               & \#4         & Ego vehicle collides with another vehicle due to insufficient steering while turning right.            \\ \midrule
			\multirow{2}{*}{Lane Invasion} & \#5         & Ego vehicle illegally crosses the solid line while passing through an intersection.                    \\
			                               & \#6         & Ego vehicle illegally crosses the solid line while turning right.                                      \\ \midrule
			\multirow{2}{*}{Stuck}         & \#7         & Ego vehicle failed to change lanes when the pedestrian in front remained stationary for too long.      \\
			                               & \#8         & Ego vehicle misjudged the state of the traffic lights, resulting in being stuck on a downhill ramp.    \\ \bottomrule
		\end{NiceTabular}
	}
\end{table}

Regarding collision violations, \tool detects 4 unique violations. We manually inspect each detected collision and categorize them into pedestrian and vehicle collisions. Below, we present three case studies (\#1, \#2, \#3 and \#4 in Table~\ref{tab:rq2}), all of which can be reproduced using \tool.

\begin{figure}[h]
	\centering
	\begin{subfigure}{0.45\textwidth}
		\centering
		\includegraphics[height=3.4cm]{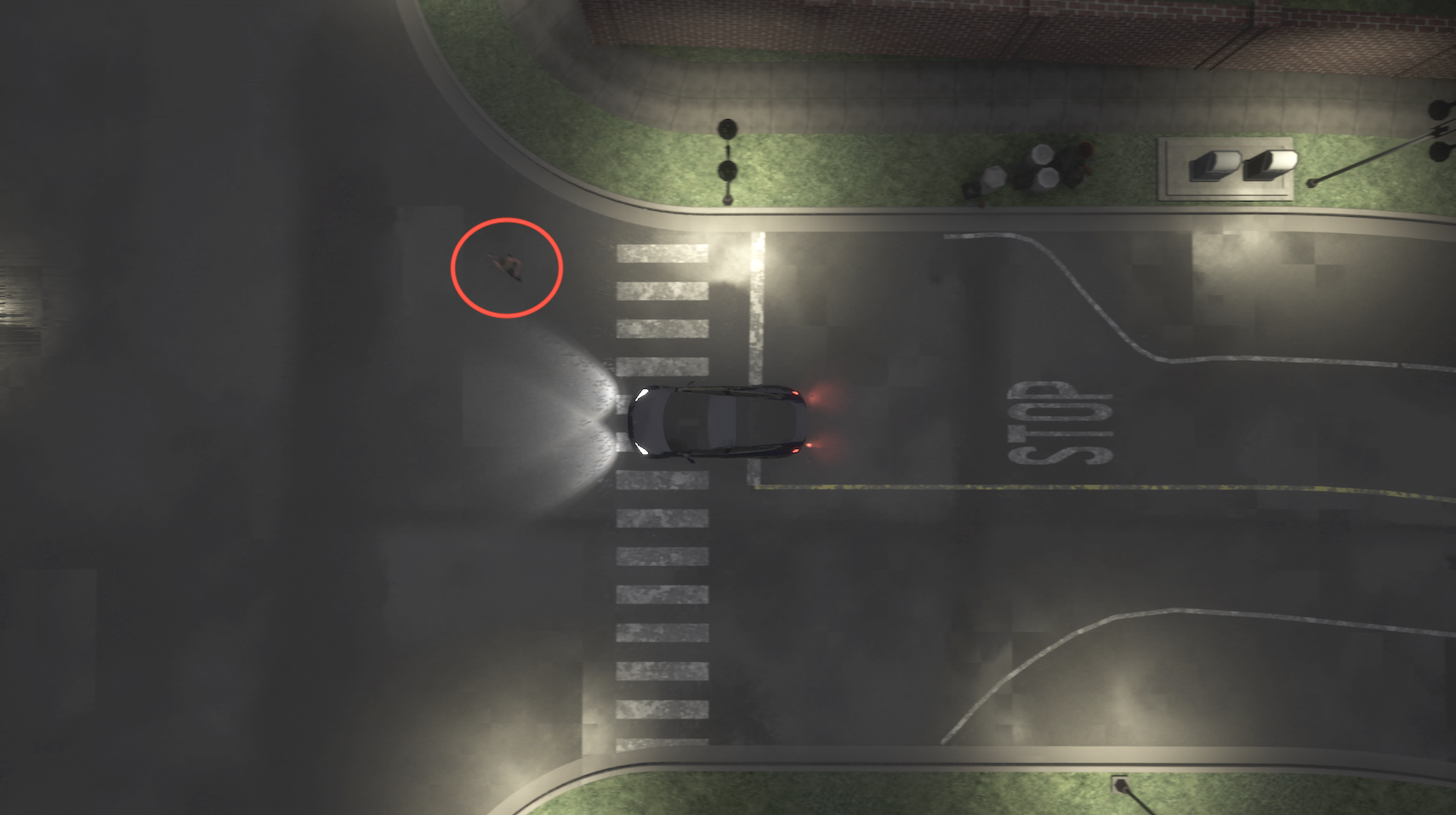}
		\caption{Bird-view of scenario at $t_1$}
	\end{subfigure}
	\hfill
	\begin{subfigure}{0.5\textwidth}
		\centering
		\includegraphics[height=3.4cm]{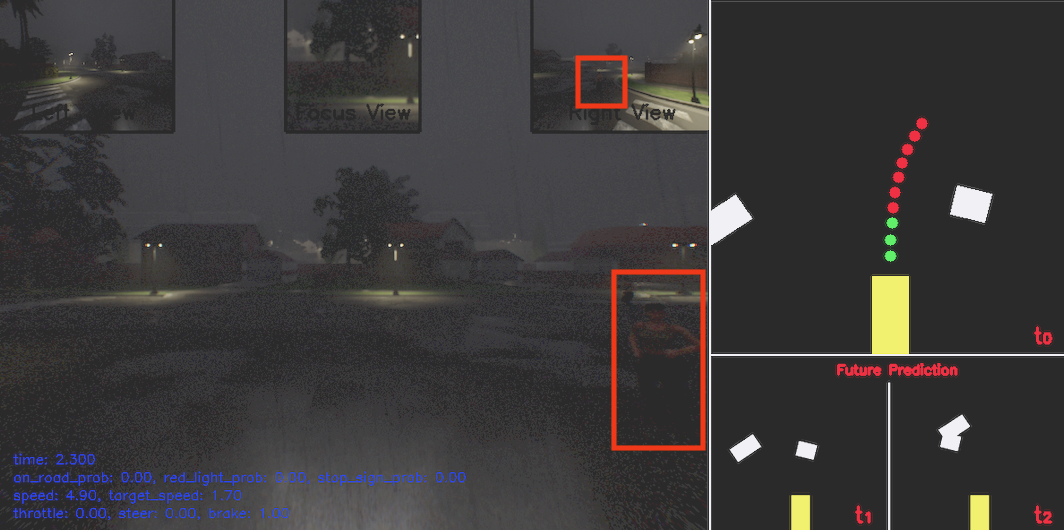}
		\caption{InterFuser controller display at $t_1$}
	\end{subfigure}
	\vspace{0.2cm}
	\begin{subfigure}{0.45\textwidth}
		\centering
		\includegraphics[height=3.4cm]{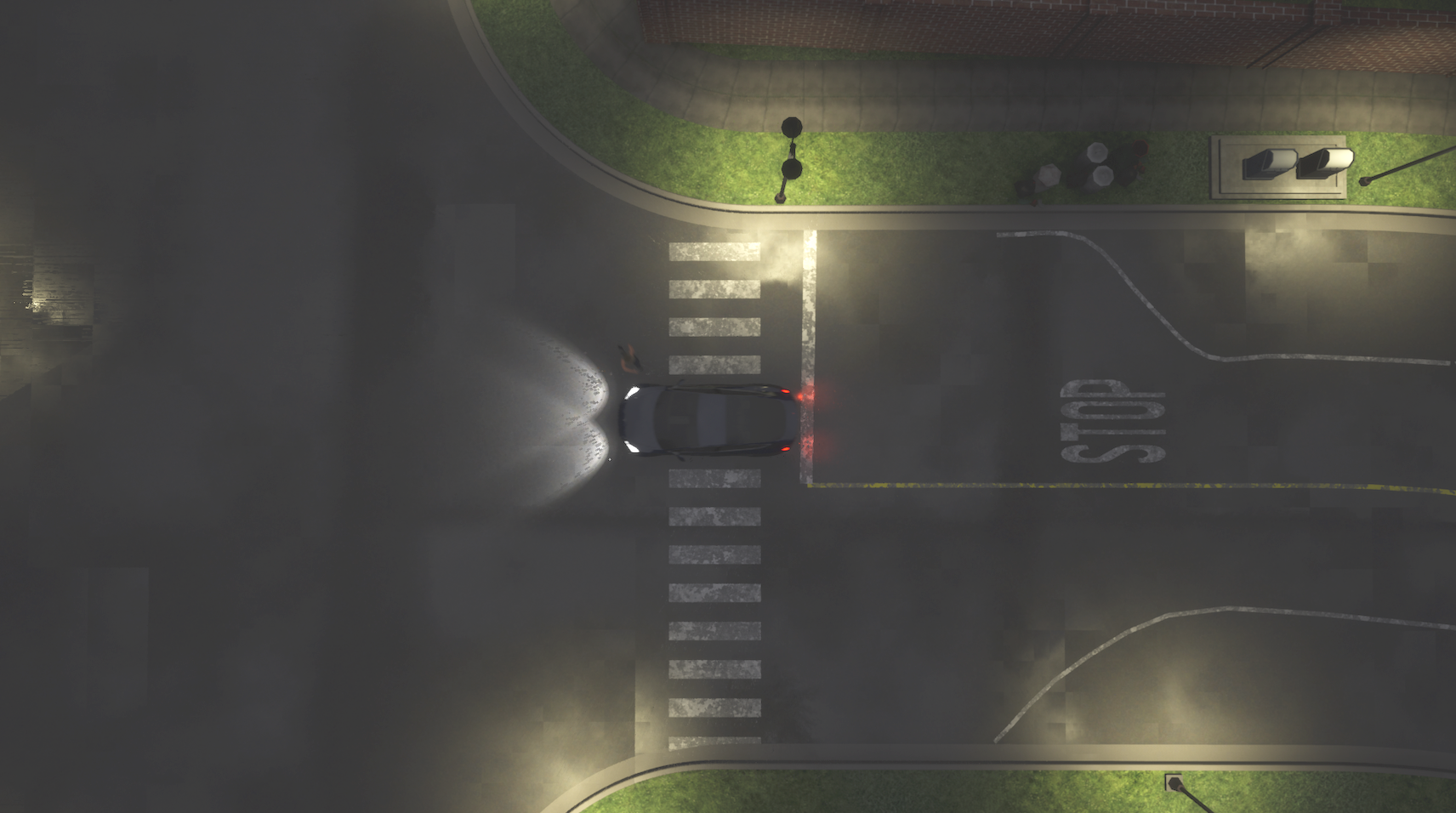}
		\caption{Bird-view of scenario at $t_2$}
	\end{subfigure}
	\hfill
	\begin{subfigure}{0.5\textwidth}
		\centering
		\includegraphics[height=3.4cm]{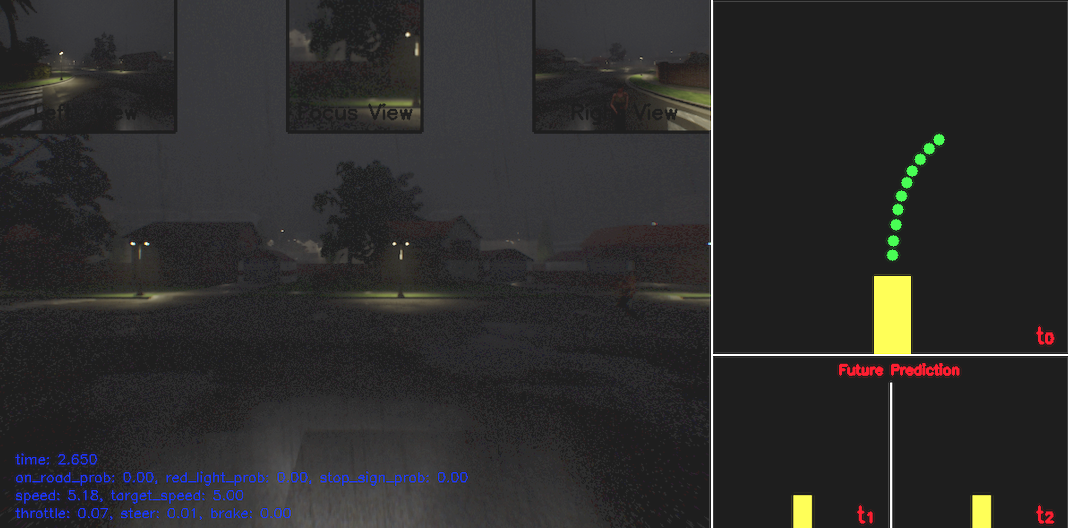}
		\caption{InterFuser controller display at $t_2$}
	\end{subfigure}
	\vspace{0.2cm}
	\begin{subfigure}{0.45\textwidth}
		\centering
		\includegraphics[height=3.4cm]{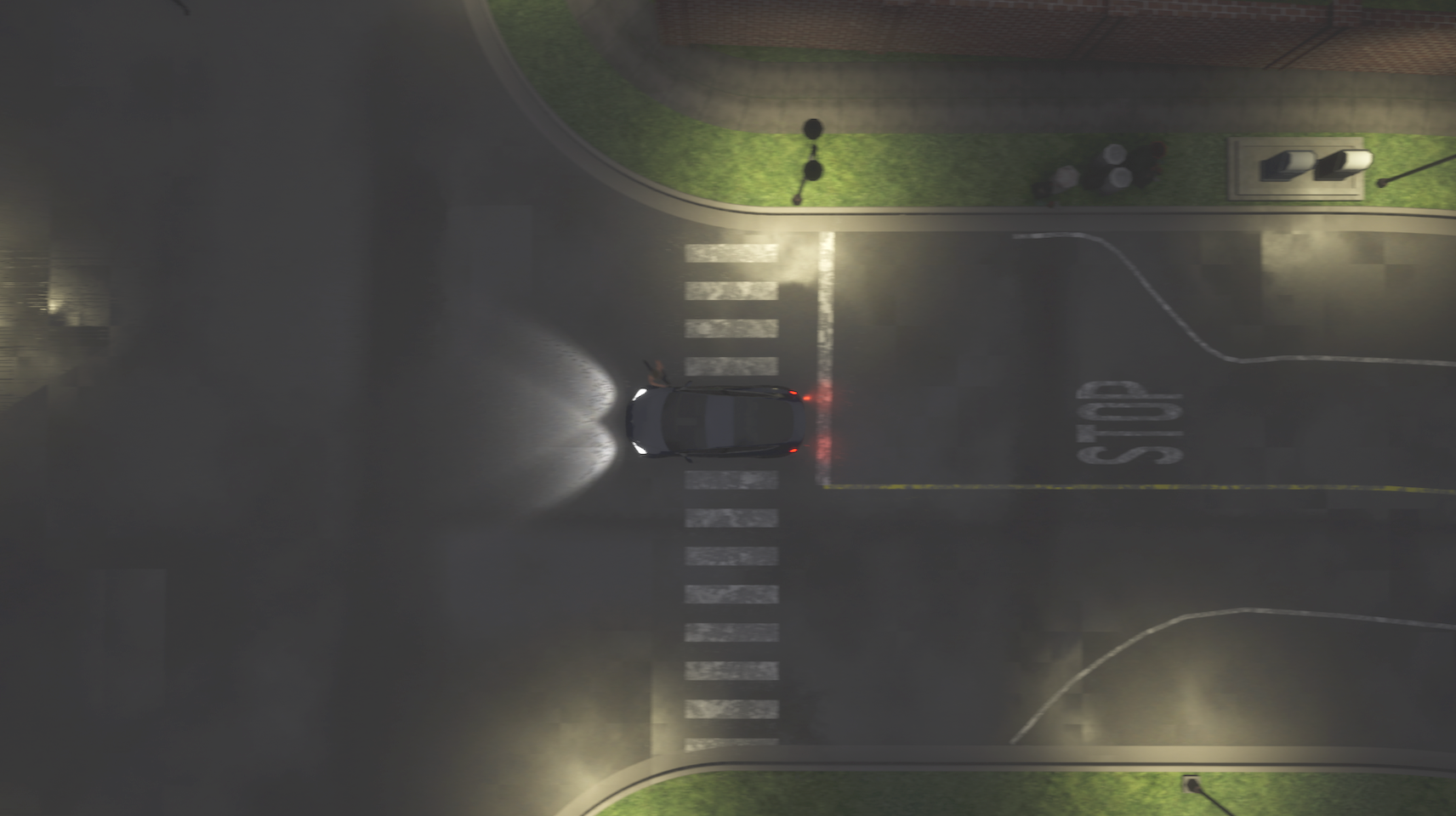}
		\caption{Bird-view of scenario at $t_3$}
	\end{subfigure}
	\hfill
	\begin{subfigure}{0.5\textwidth}
		\centering
		\includegraphics[height=3.4cm]{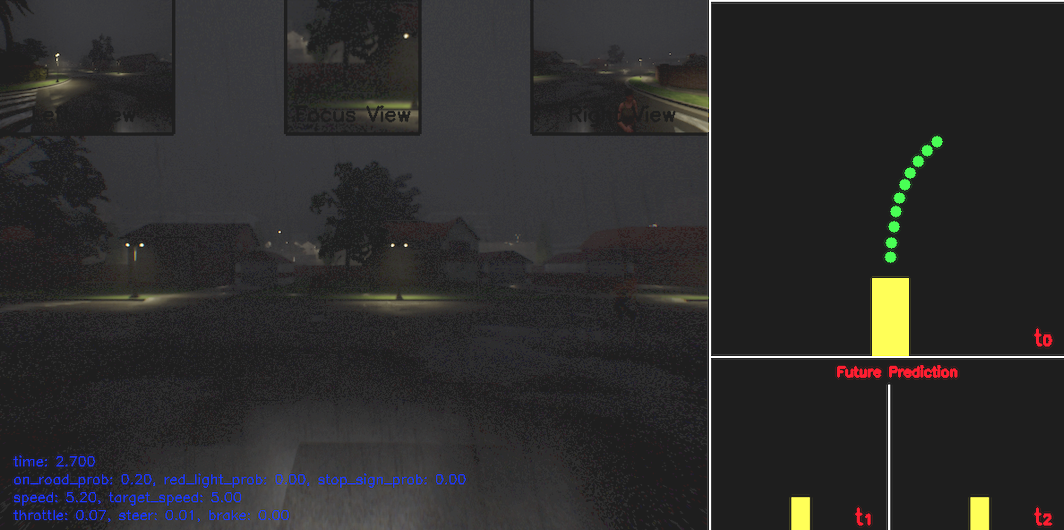}
		\caption{InterFuser controller display at $t_3$}
	\end{subfigure}

	\caption{Pedestrian Collision under Low-Light Conditions (\#1)}
	\label{fig:rq2_case_walker}
\end{figure}

\textbf{Case Study\#1: Pedestrian Collision by InterFuser.}
As shown in Figure~\ref{fig:rq2_case_walker}, the collision occurs at night under low-light conditions. Although \tool ensures that vehicle headlights are activated when the sun altitude falls below 90 degrees, the driving conditions remain more challenging than in daylight. InterFuser successfully detects a moving object at timestamp $t_1$ (Figure~\ref{fig:rq2_case_walker}(b)), but it fails to recognize the pedestrian at timestamps $t_2$ and $t_3$. From the control signals generated by InterFuser (highlighted in blue in the controller display at the bottom-right), we observe that the ego vehicle does not apply any braking at timestamps $t_2$ and $t_3$. Although the pedestrian's behavior contributed to the collision, InterFuser's failure to take appropriate actions (\ie braking) also makes it partially responsible for the accident.

\begin{figure}[h]
	\centering
	\begin{subfigure}{0.30\textwidth}
		\centering
		\includegraphics[height=2.3cm]{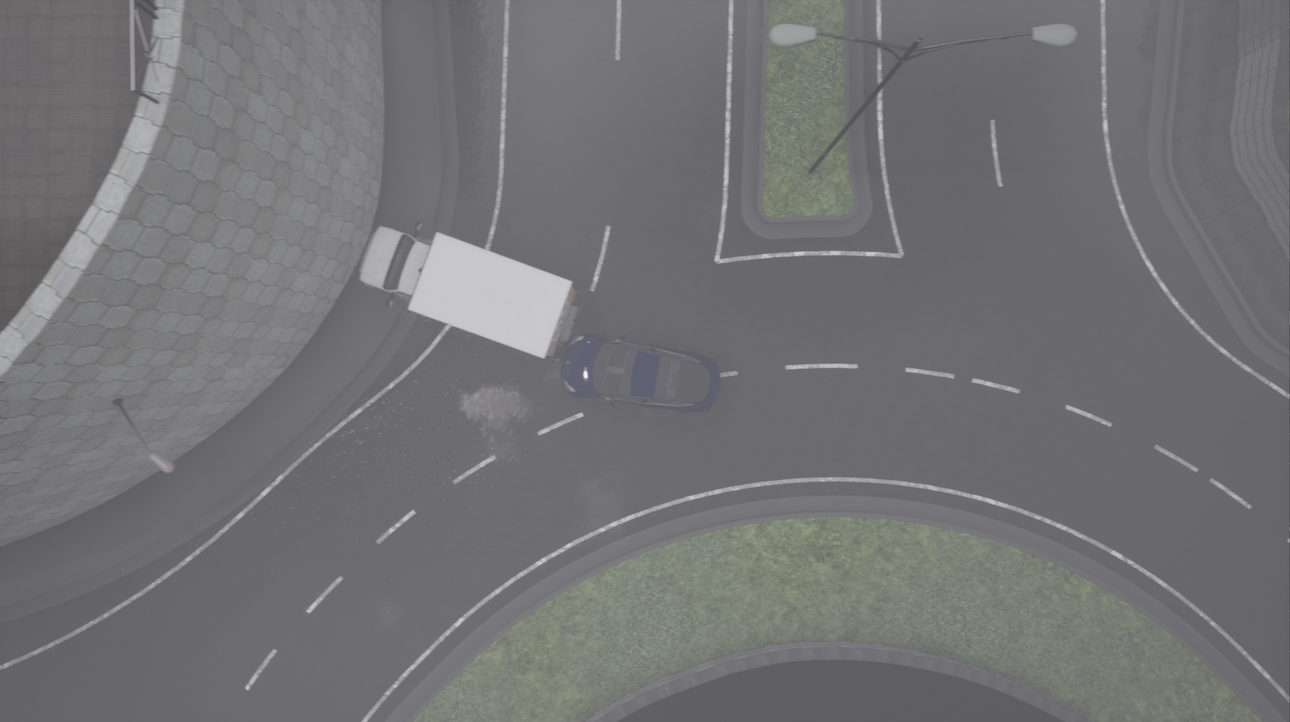}
		\caption{Bird-view of scenario}
	\end{subfigure}
	\hfill
	\begin{subfigure}{0.30\textwidth}
		\centering
		\includegraphics[height=2.3cm]{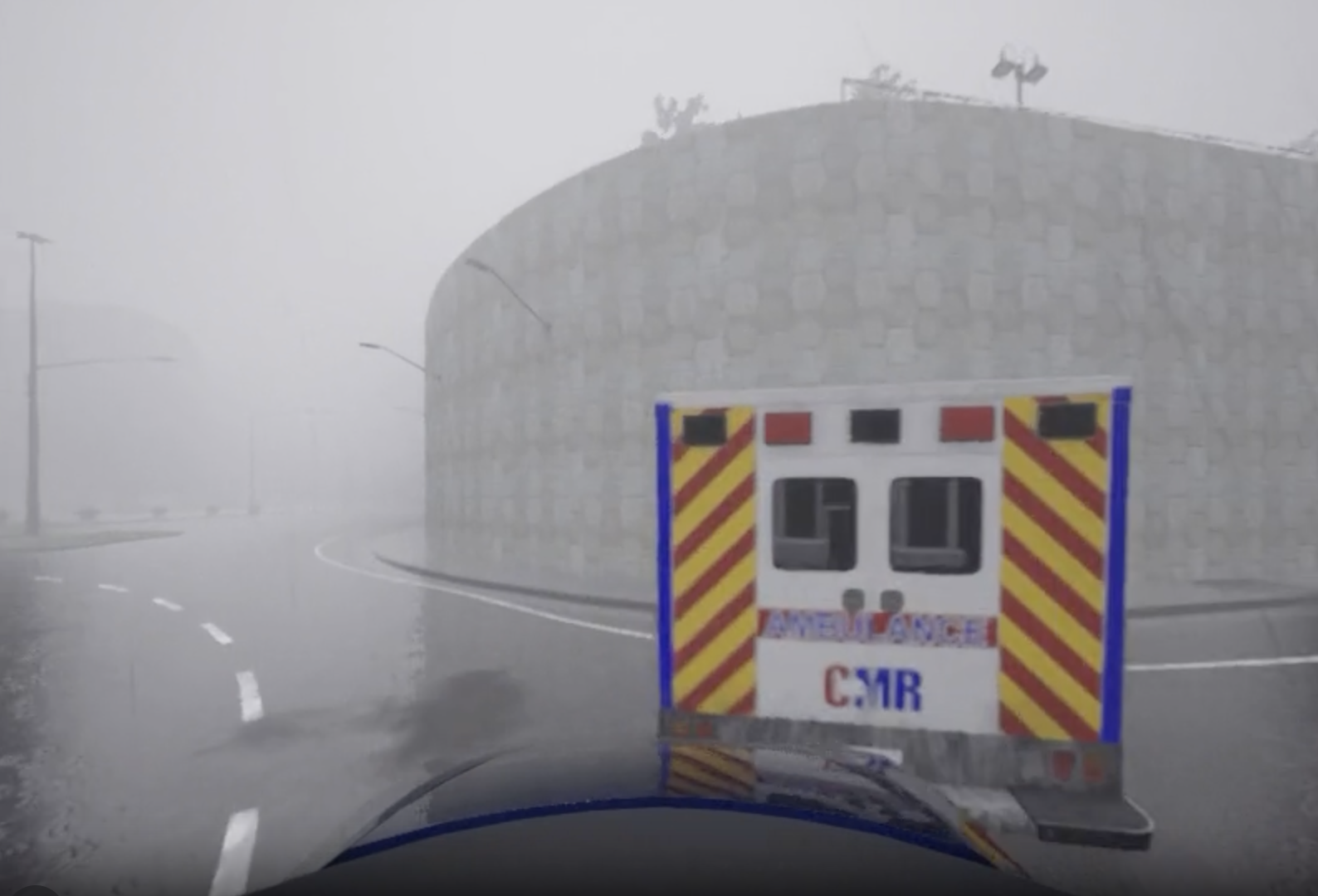}
		\caption{Front-view of scenario}
	\end{subfigure}
	\hfill
	\begin{subfigure}{0.35\textwidth}
		\centering
		\includegraphics[height=2.3cm]{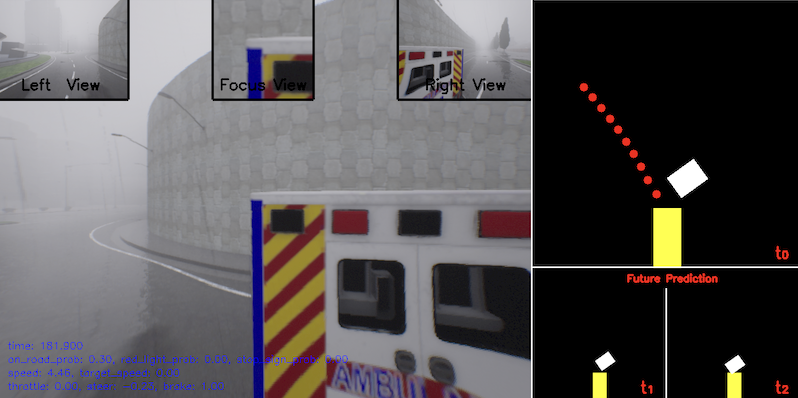}
		\caption{InterFuser controller display}
	\end{subfigure}
	\caption{Vehicle Collision during Lane Change at the Roundabout (\#2)}
	\label{fig:rq2_case_car}
\end{figure}

\begin{figure}[h]
	\centering
	\begin{subfigure}{0.45\textwidth}
		\centering
		\includegraphics[height=3.4cm]{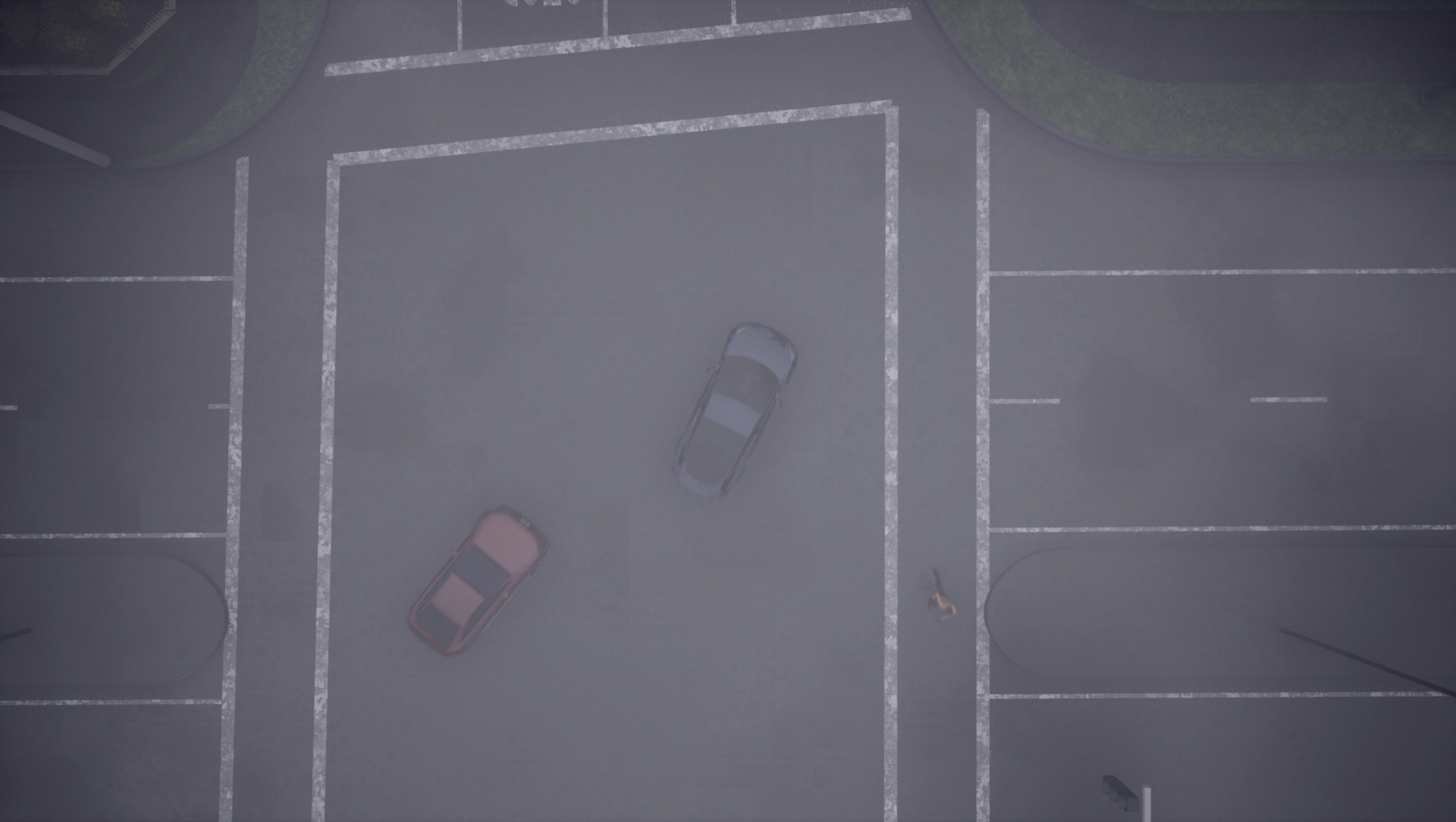}
		\caption{Bird-view of scenario at $t_1$}
	\end{subfigure}
	\hfill
	\begin{subfigure}{0.52\textwidth}
		\centering
		\includegraphics[height=3.4cm]{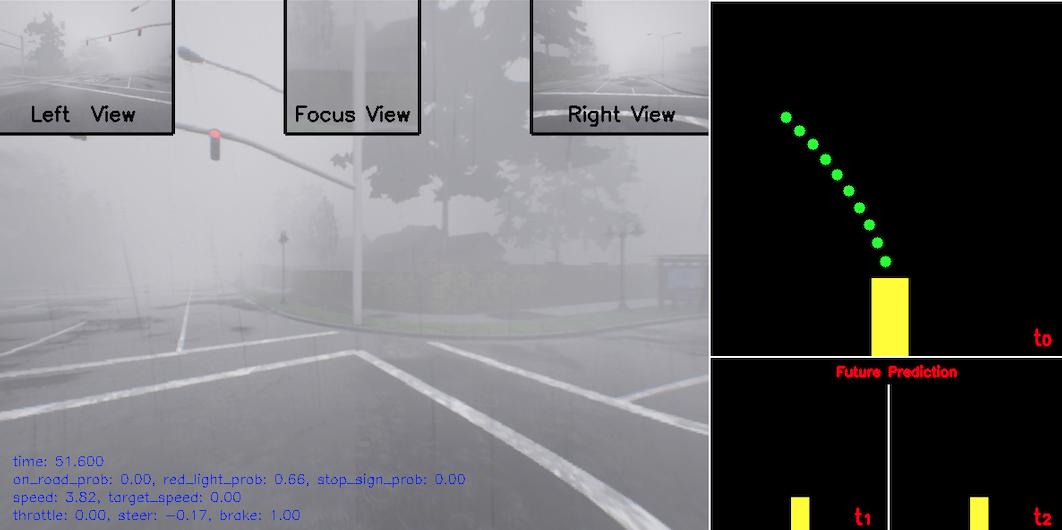}
		\caption{InterFuser controller display at $t_1$}
	\end{subfigure}
	\vspace{0.2cm}
	\begin{subfigure}{0.45\textwidth}
		\centering
		\includegraphics[height=3.4cm]{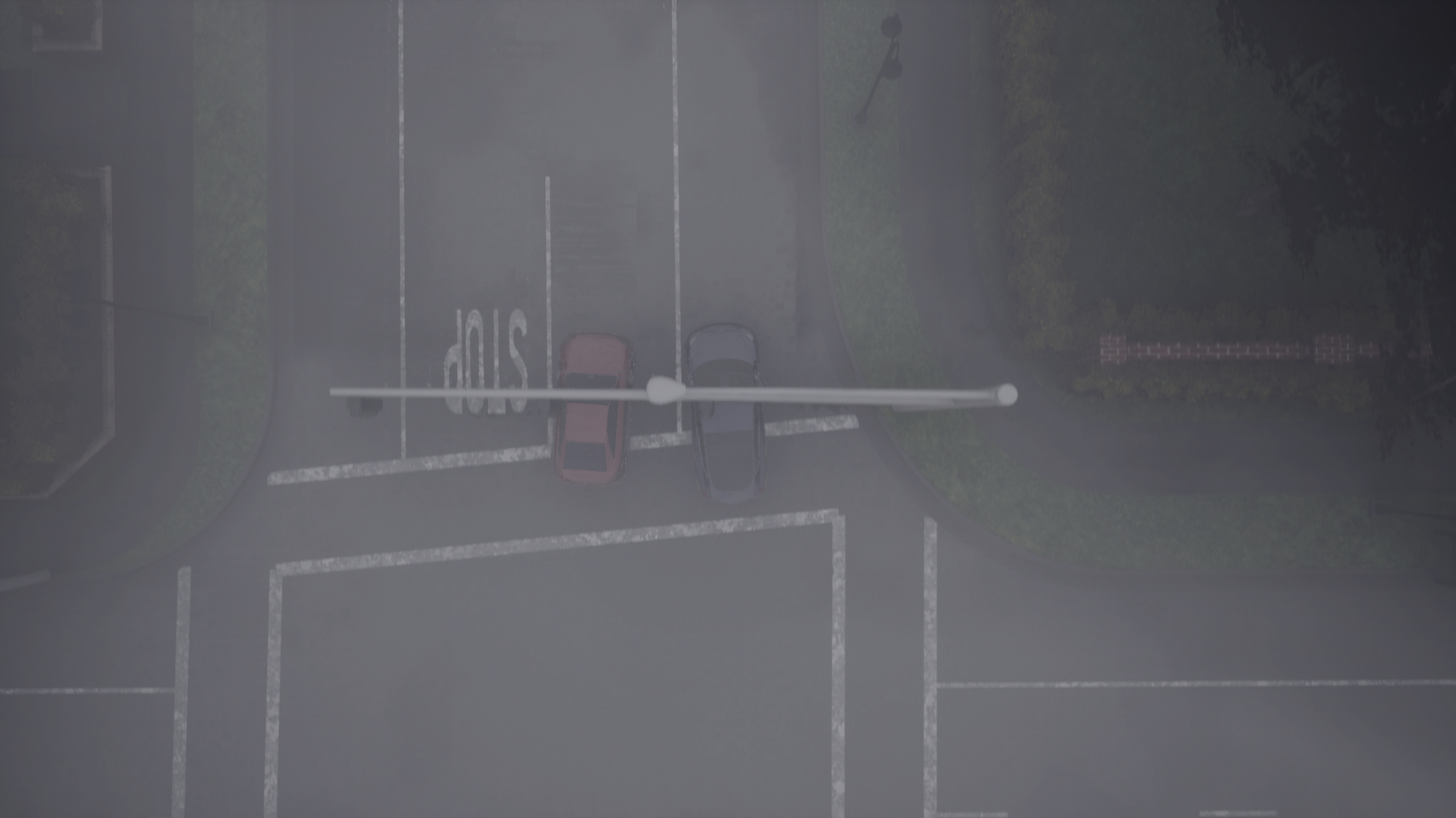}
		\caption{Bird-view of scenario at $t_2$}
	\end{subfigure}
	\hfill
	\begin{subfigure}{0.52\textwidth}
		\centering
		\includegraphics[height=3.4cm]{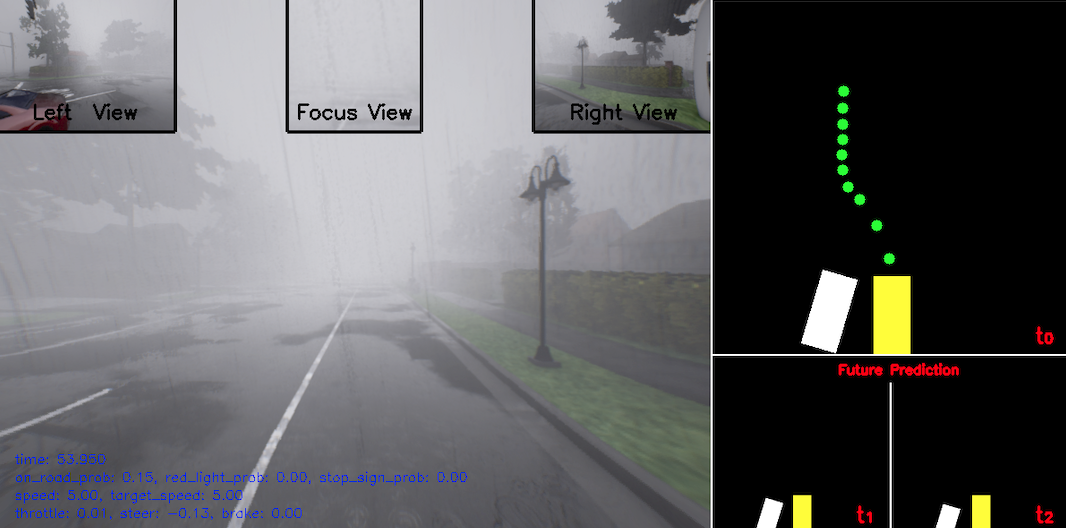}
		\caption{InterFuser controller display at $t_2$}
	\end{subfigure}
	\vspace{0.2cm}
	\begin{subfigure}{0.45\textwidth}
		\centering
		\includegraphics[height=3.4cm]{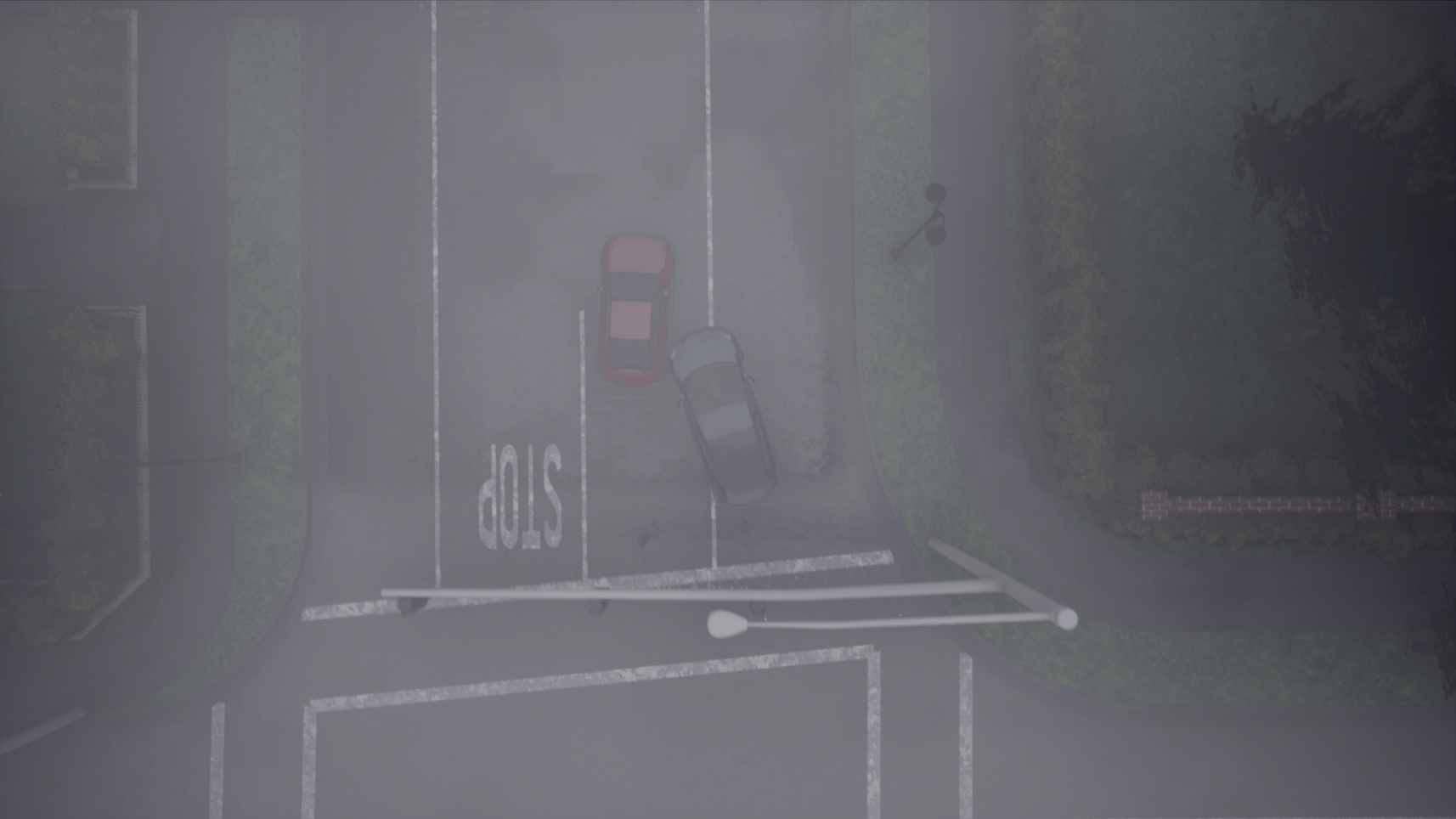}
		\caption{Bird-view of scenario at $t_3$}
	\end{subfigure}
	\hfill
	\begin{subfigure}{0.52\textwidth}
		\centering
		\includegraphics[height=3.4cm]{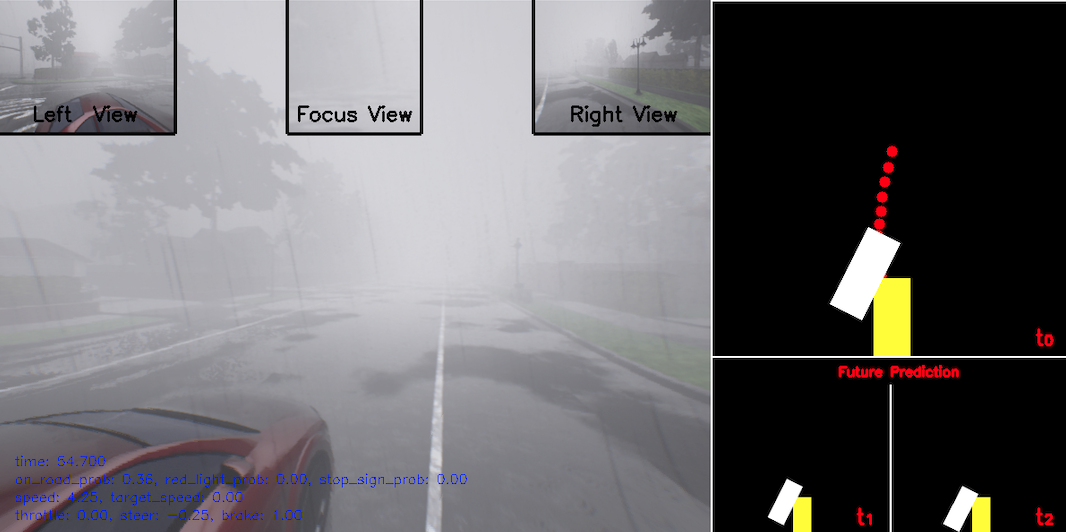}
		\caption{InterFuser controller display at $t_3$}
	\end{subfigure}
	\caption{Vehicle Collision during Lane Change at the Intersection (\#3)}
	\label{fig:rq2_case_car_intersection}
\end{figure}

\textbf{Case Study\#2, \#3 and \#4: Vehicle Collision by InterFuser.}
As shown in Figure~\ref{fig:rq2_case_car}, the ego vehicle rear-collides with a van when changing lanes to exit the roundabout. At the time, the van is stationary in the ego vehicle's path. From Figure~\ref{fig:rq2_case_car}(c), we can see that while InterFuser successfully detects an object to the right of the ego vehicle, it misjudges the size of the obstacle. This misjudgment results in insufficient left steering by the ego vehicle, ultimately causing the collision. Figure~\ref{fig:rq2_case_car_intersection}

Figure~\ref{fig:rq2_case_car_intersection} illustrates a collision caused by an improper lane change. At $t_1$, two vehicles are turning left at an intersection, with the front vehicle controlled by InterFuser. Then, at $t_2$, the ego vehicle merges into the side lane while a red car is driving alongside it on the left. Although InterFuser identified the red car, it failed to yield and persisted in making the lane change despite the situation. At $t_3$, a side collision occurred due to an incorrect estimation of the red vehicle’s speed.

\begin{figure}[h]
	\centering
	\begin{subfigure}{0.3\textwidth}
		\centering
		\includegraphics[height=2.3cm]{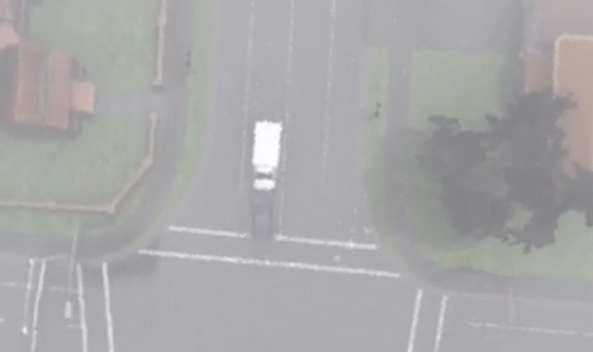}
		\caption{Bird-view of scenario}
	\end{subfigure}
	\hfill
	\begin{subfigure}{0.3\textwidth}
		\centering
		\includegraphics[height=2.3cm]{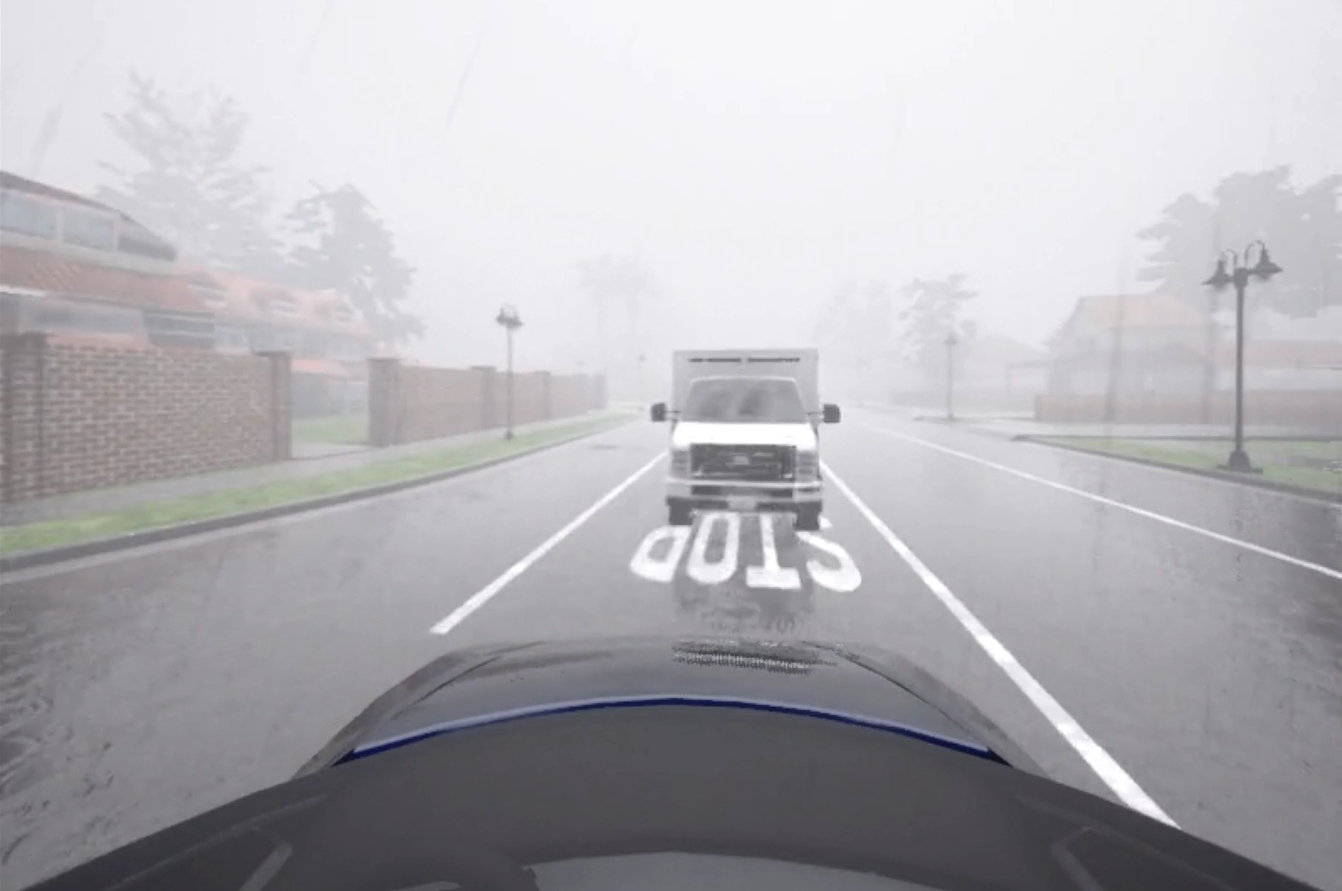}
		\caption{Front-view of scenario}
	\end{subfigure}
	\hfill
	\begin{subfigure}{0.35\textwidth}
		\centering
		\includegraphics[height=2.3cm]{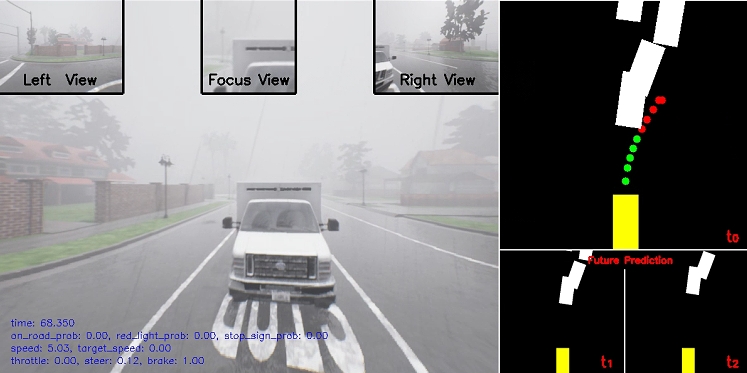}
		\caption{InterFuser controller display}
	\end{subfigure}
	\caption{Vehicle Collision during Turn Right (\#4)}
	\label{fig:rq2_case_car_2}
\end{figure}

Figure~\ref{fig:rq2_case_car_2} shows a more severe frontal collision. The ego vehicle, turning right at an intersection with insufficient steering, drives into the opposite lane. Unfortunately, a white truck is approaching from the opposite direction. Although the truck is braking, and InterFuser immediately outputs a brake signal (as shown in Figure~\ref{fig:rq2_case_car_2}(c)), trying to avoid the collision by stopping in front of the truck. However, the ego vehicle's speed is too high to stop in time, and ultimately, the two vehicles collide head-on.

\begin{tcolorbox}[]
	{ \textbf{Answer to RQ2: } \tool reveals 32, 27, and 9 more unique violations than AV-Fuzzer, DriveFuzz, and TM-Fuzzer, demonstrating its effectiveness in generating high-quality scenarios for simulation-based testing.}
\end{tcolorbox}


\subsubsection{RQ3: Diversity of Scenarios}

The main goal of \tool is to generate more diverse and interactive driving scenarios. One of the metrics used to evaluate diversity is ego vehicle trajectory coverage on the map~\cite{hu2021coverage}. We mark the waypoints on the Town03 map at 5-meter intervals and define trajectory coverage as the number of waypoints covered by the ego vehicle’s route divided by the total number of waypoints on the map.

\begin{figure}[h]
	\centering

	\begin{subfigure}{0.45\textwidth} 
		\centering
		\includegraphics[width=\textwidth]{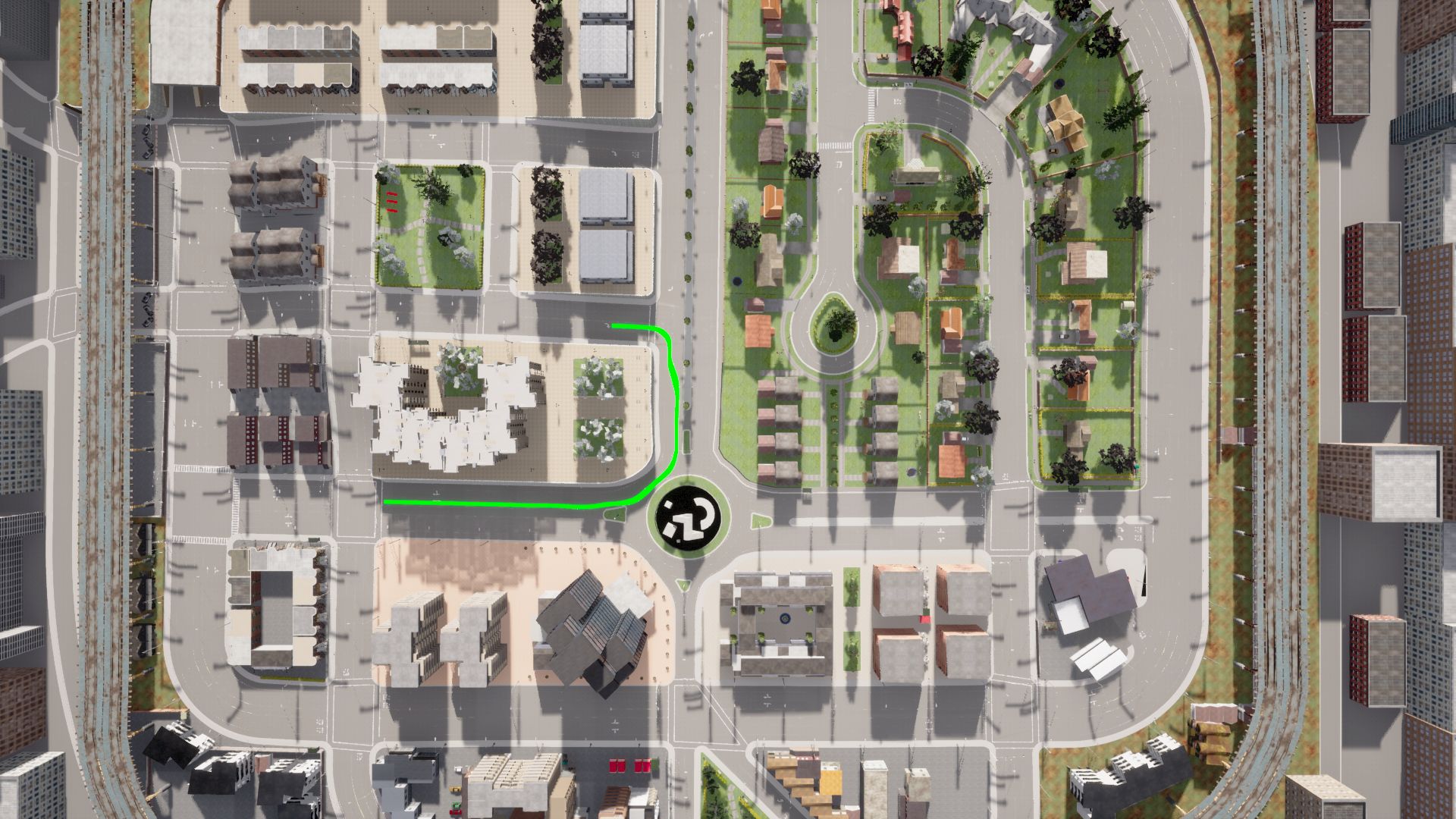}
		\caption{AV-Fuzzer (3.04\%)}
		\label{fig:sub1}
	\end{subfigure}
	\hfill 
	\begin{subfigure}{0.45\textwidth} 
		\centering
		\includegraphics[width=\textwidth]{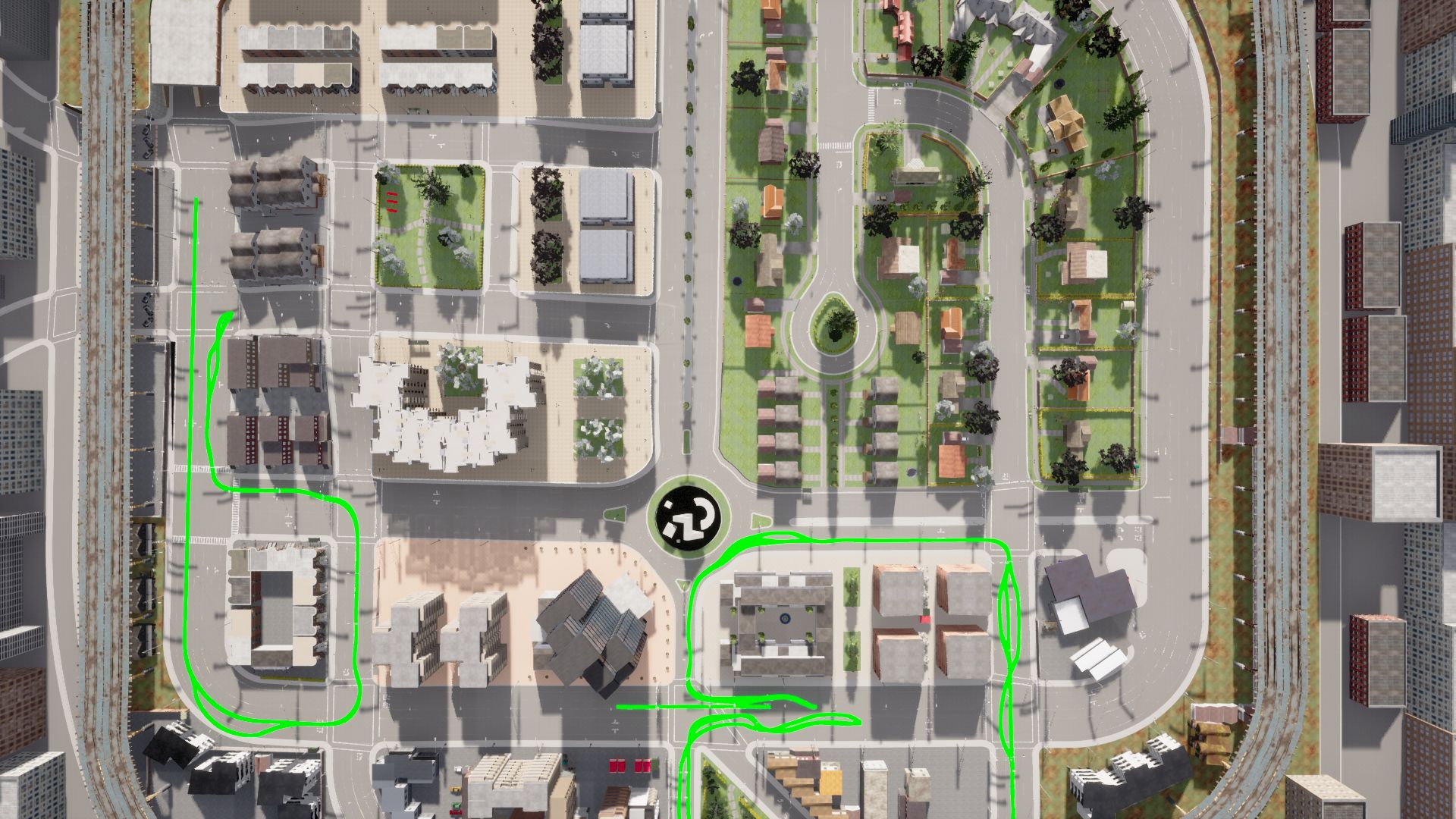} 
		\caption{DriveFuzz (13.85\%)}
		\label{fig:sub2}
	\end{subfigure}

	\vspace{0.2cm}

	\begin{subfigure}{0.45\textwidth}
		\centering
		\includegraphics[width=\textwidth]{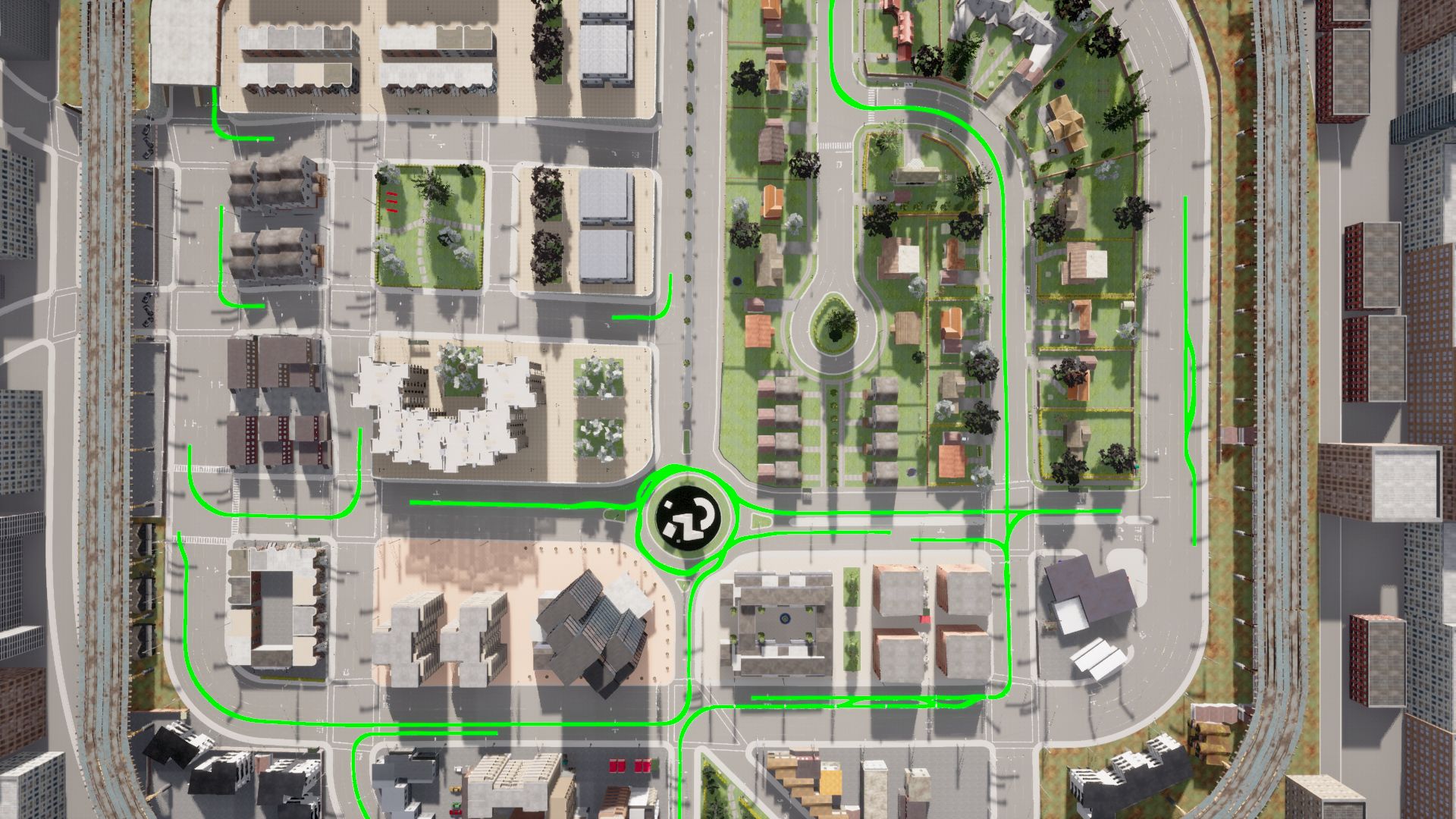} 
		\caption{TM-Fuzzer (24.88\%)}
		\label{fig:sub3}
	\end{subfigure}
	\hfill 
	\begin{subfigure}{0.45\textwidth} 
		\centering
		\includegraphics[width=\textwidth]{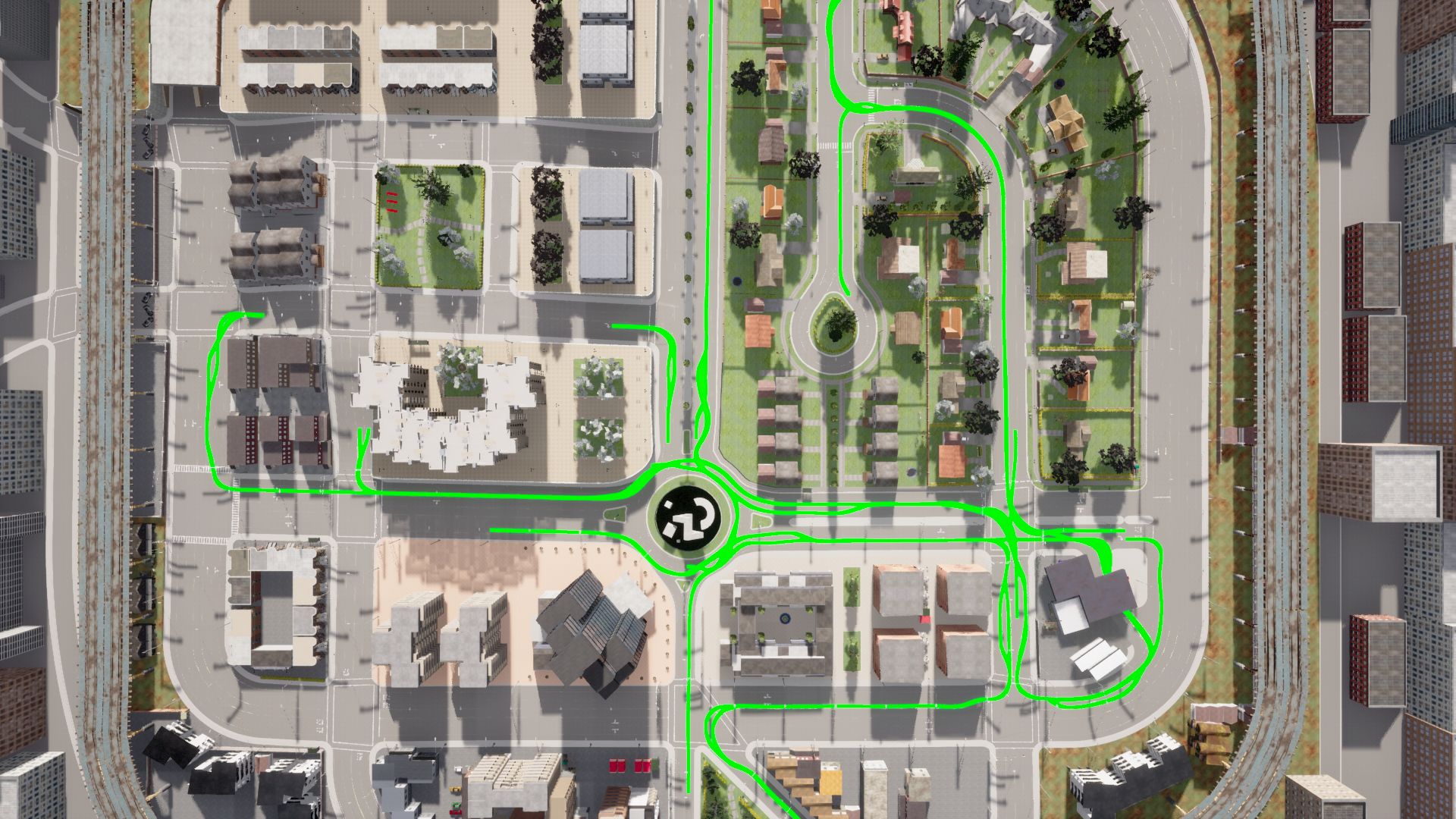}
		\caption{SimADFuzz-R+R (19.65\%)}
		\label{fig:sub4}
	\end{subfigure}

	\vspace{0.2cm}

	\begin{subfigure}{0.45\textwidth}
		\centering
		\includegraphics[width=\textwidth]{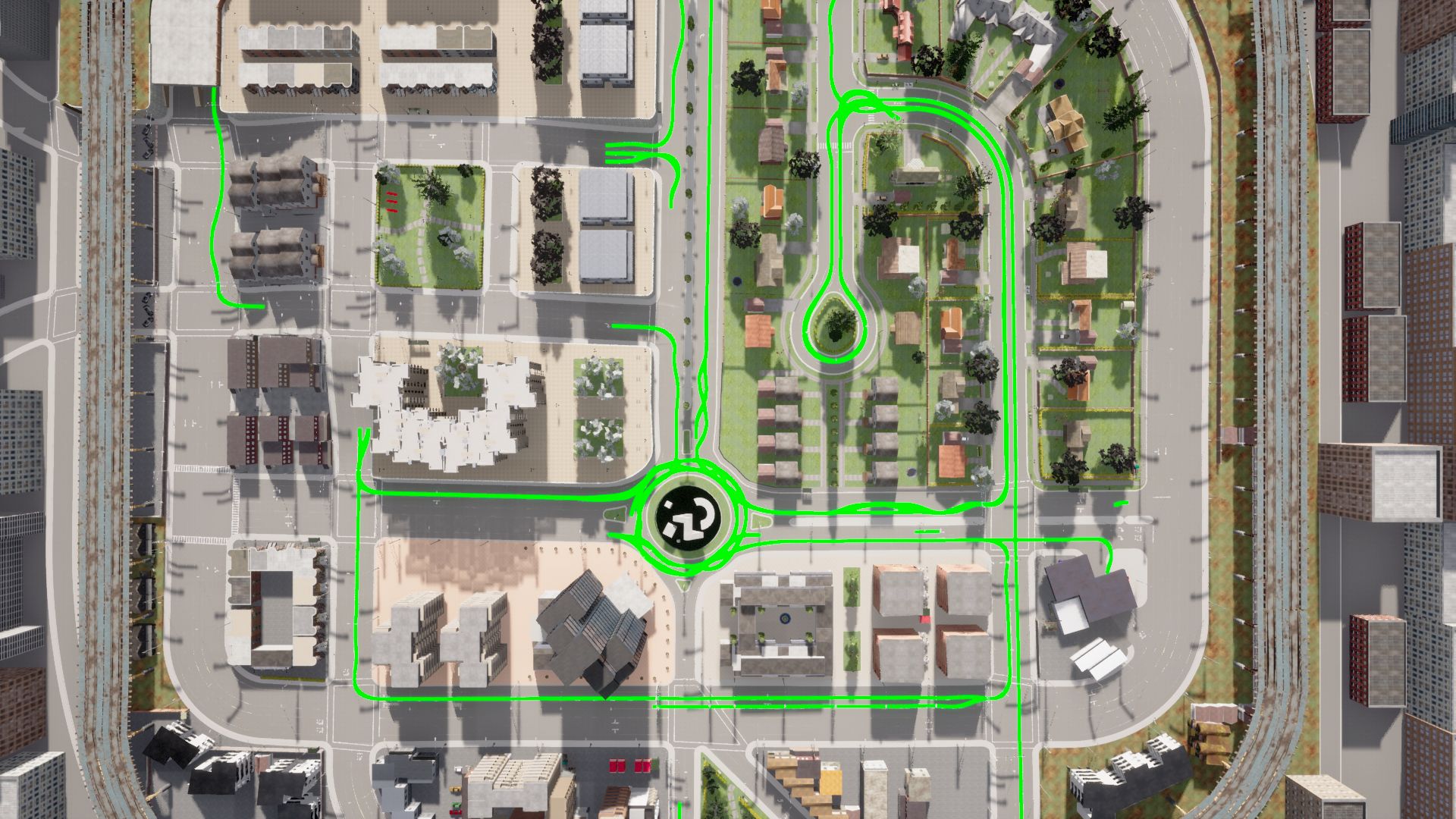} 
		\caption{SimADFuzz-VS+R (29.32\%)}
		\label{fig:sub5}
	\end{subfigure}
	\hfill
	\begin{subfigure}{0.45\textwidth} 
		\centering
		\includegraphics[width=\textwidth]{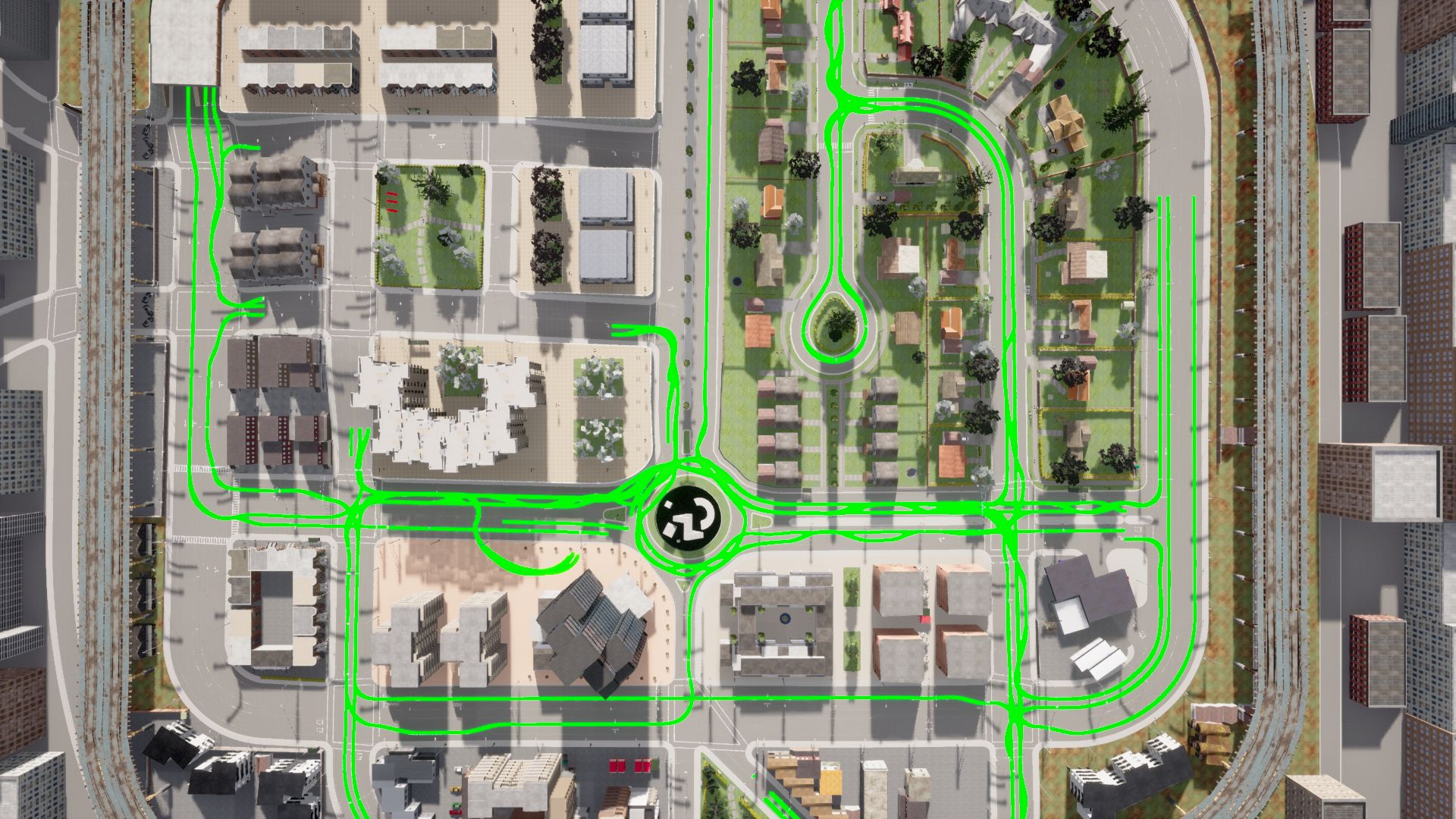}
		\caption{SimADFuzz-VS+D (61.25\%)}
		\label{fig:sub6}
	\end{subfigure}

	\caption{Trajectory Coverage of Fuzzers in Town03}
	\label{fig:rq4}
\end{figure}

As shown in Figure~\ref{fig:rq4}, \tool (\ie VS+D) achieved 61.25\% trajectory coverage, significantly improving upon AV-Fuzzer (3.04\%), which does not mutate the ego vehicle’s route but only the NPCs, and DriveFuzz (13.85\%), which mutates the ego vehicle’s route only after the cycle and mutation process are completed. TM-Fuzzer, which aims to increase interactions by dynamically controlling traffic flow, achieved a trajectory coverage of 24.88\%.

Recall that \tool detects more violations than other fuzzers, which indicates that \tool not only increases the diversity of the ego vehicle's trajectory across the map but also improves the quality of the scenarios for detecting risky behaviors.

\begin{tcolorbox}[]
	{\textbf{Answer to RQ3:} The strategies used in \tool enhance the coverage of the ego vehicle’s trajectory on the map, thereby improving the diversity of the generated scenarios.}
\end{tcolorbox}


\subsection{Threats to Validity}

\subsubsection{Internal Validity}
One potential threat to internal validity is the implementation of \tool. We developed \tool based on the DEAP and CARLA simulator and extended it by (1) enhancing the simulation information collector through analysis of actions determined by InterFuser, (2) integrating a model-based fitness evaluation using a deep neural network model and adapting the fitness score to the NSGA-2 algorithm provided by DEAP, and (3) customizing the scenario mutation procedure by introducing a distance-guided mutation strategy. Although the implementation of \tool has undergone peer review, there may still be issues that could affect the experimental results. To mitigate this risk, we have made the source code of \tool publicly available, allowing the community to reproduce the results and validate the implementation.

\subsubsection{External Validity}
A potential threat to external validity is that we evaluate the effectiveness of \tool only on InterFuser, which may limit the generalizability of the results to other ADS. However, it is important to note that InterFuser has achieved top-performing results on the CARLA Leaderboard. In future work, we plan to extend \tool to other ADS (\eg Apollo and Autoware) and validate its effectiveness across a broader range of systems.
\subsubsection{Construct Validity}
The primary threat to construct validity lies in the metrics used to evaluate the effectiveness of \tool. In our experiment, we focus solely on the number of unique violations detected by each fuzzer. The violation oracle used in \tool, its variants, and other fuzzers are identical. Therefore, comparing the number of unique violations provides a fair assessment of the performance of \tool. Furthermore, the number of unique violations is a widely accepted metric in the evaluation of fuzz testing for ADS~\cite{cheng2023behavexplor,Doppel}, and it effectively reflects the capability of \tool in detecting violations within ADS.
\section{Related Work}
\label{sect:related}

\tool leverages fuzzing techniques for simulation-based testing, and proposes novel scenario selection and mutation strategies to generate offspring scenarios. This section reviews the related work on simulation-based testing for ADS and optimization techniques for fuzz testing.

\subsection{Simulation-based Testing for ADS}

Simulation-based testing presents a viable and efficient alternative, enabling the exploration of a broad spectrum of scenarios and environments in a controlled and safe setting~\cite{nalic2020scenario}. However, as the complexity of simulated environments increases, the number of configurable and variable elements also grows, leading to a vast space of possible scenarios. Therefore, simulation-based testing methods for ADS focus on how to generate critical scenarios.

Wang et al.~\cite{AdvSim} presented AdvSim, a framework that generates safety-critical scenarios by adversarially altering the trajectories of actors within traffic scenarios to test the LiDAR-based ADS. Gambi et al.~\cite{Asfault} presented ASFault, which creates road networks to simulate driving scenarios and employs a genetic algorithm to generate tests aimed at exposing unsafe behaviors in lane-keeping systems.

Moreover, Sun et al.~\cite{sun2022lawbreaker} and Zhang et al.~\cite{zhang2023testing} expanded the oracle of violations beyond collisions to traffic laws. They presented LawBreaker and GFlowNet, respectively, which assess the "distance" between the behavior of ego vehicles and traffic laws using signal temporal logic~\cite{maler2004monitoring}, thereby generating scenarios that more closely resemble violations. Huai et al.~\cite{Doppel} presented Doppel, which replaces non-intelligent agents in the simulation with the ego vehicle, ensuring that all violations are triggered by the ego vehicle and reducing false positives. Lu et al.~\cite{9712397} proposed DeepCollision based on the Deep Q-Learning (DQN) algorithm~\cite{mnih2015human}, which calculates the safety distance and current distance to evaluate the collision probability, and designs reward functions to make the DQN output the optimal action, thereby increasing the possibility of detecting collision behaviors. Tian et al. proposed MOSAT~\cite{MOSAT} and CRISCO~\cite{CRISCO}. MOSAT represents vehicle driving behavior as a gene sequence composed of atomic driving maneuvers. It considers fitness metrics such as the estimation time to collision (ETTC) and leverage NSGA-2 to select scenarios, then the maneuver sequence are mutated to generate offspring driving scenarios. CRISCO extracts influential behavior patterns from historical traffic accidents, and assigns participants to move along specified trajectories during scenario generation. Additionally, CRISCO leverages ETTC to evaluate the criticality of driving scenarios, selecting more critical scenarios to challenge the ADS.

Haq et al.~\cite{Haq0B22} focus on DNN-enabled systems and propose an online testing method that utilizes MOSA/FITEST as the multi-objective search algorithm to guide the generation of test cases. In contrast, \tool employs the NSGA-2 algorithm, which is orthogonal to MOSA/FITEST. While both aim to optimize multiple objectives, they do so through different methodologies and strategies, offering unique advantages in different contexts, such as exploration vs. exploitation trade-offs.

AV-Fuzzer~\cite{AV-Fuzzer}, AutoFuzz~\cite{AutoFuzz}, DriveFuzz~\cite{DriveFuzz}, TM-Fuzzer~\cite{lin2024tm}, scenoRITA~\cite{scenoRITA} and ScenarioFuzz~\cite{wang2024dance} are representative methods that utilize fuzz testing techniques to generate scenarios for autonomous driving simulations. These fuzzers adapt their scenario generation strategies based on simulation feedback. For instance, AV-Fuzzer retains scenarios with higher potential safety risks when interacting with other actors. AutoFuzz evaluates objectives such as the ego vehicle's speed during collisions and the minimum distance to other actors. Similarly, DriveFuzz incorporates metrics such as hard braking, steering, and minimum distances as driving quality scores, selecting the scenarios with the highest fitness to generate offspring. TM-Fuzzer utilizes real-time traffic management and diversity analysis to serach and generate critical and unique scenarios. scenoRITA optimize mutation strategies force on obstacles states (\eg vehicles, pedestrians and bikes) and proposed a technique to eliminate duplicate scenarios to enhance testing efficiency.
The most similar work with \tool is ScenarioFuzz, which builds a graph neural network to predict and filter out high-risk scenarios, while we leverage Transformer model to embed the scenarios by considering the temporal features of the ego vehicle and NPC vehicles. Note that \tool selects scenarios based on multi-objective search algorithm NSGA-2, so it is convenient to adapt other fitness (such as confidence level proposed by ScenarioFuzz) to further refine our method.

\subsection{Optimization Techniques for Fuzz Testing}

Fuzz testing is a dynamic software analysis and testing technique that employs random inputs as test cases, which are then executed within the programs under test (PUT)~\cite{liang2018fuzzing}. The quality of test cases is pivotal to the efficacy of fuzz testing; high-quality test cases facilitate the exploration of more diverse execution paths within the program, while low-quality test cases may lead to inefficient resource allocation and reduced testing effectiveness. Thus, recent work focused on optimizing seed selection and mutation strategies to generate high-quality test cases.

For seed selection, Memlock~\cite{wen2020memlock} prioritizes seeds based on coverage and memory consumption to uncover vulnerabilities related to memory usage in PUT. Truzz~\cite{zhang2022path} favors seeds that generate new edge coverages, thereby improving the code coverage. K-Scheduler~\cite{she2022effective} assigns different probabilities for selection based on the centrality of each seed. Cerebro~\cite{li2019cerebro} calculates a comprehensive score for seeds based on the complexity of unexplored code near the execution path. AFLSmart~\cite{pham2019smart} computes the effectiveness ratio of seeds, with higher effectiveness seeds being allocated more resources, thus increasing the likelihood of generating effective offspring test cases.

For seed mutation, several works leverage machine learning models to optimize mutation strategies. Neuzz~\cite{she2019neuzz} employs a neural network model to smoothen complex PUT, establishing a relationship between test case byte sequences and branch coverage, identifying key byte positions through gradients for targeted mutation. MTFuzz~\cite{she2020mtfuzz} uses a Multi-Task neural network to learn a compact embedding of the input space for multiple related tasks, guiding the mutation process by focusing on high-gradient areas of the embedding. Wu et al.~\cite{wu2022evaluating} proposed PreFuzz, which enhances gradient guidance through a resource-efficient edge selection mechanism and a probabilistic byte selection mechanism to improve mutation effectiveness.

Existing seed selection and mutation optimization strategies cannot be directly applied to fuzz testing for ADS, because the scenarios are well-defined structured data compared to byte sequences. Inspired by SmarTest~\cite{smartest}, which leverages language models to prioritize and generate transaction sequences (\ie test cases in the context of smart contracts), {\tool} employs a model-based fitness score evaluation method to optimize scenario selection. Furthermore, by measuring the distance between the ego vehicle and other traffic participants, the mutation strategy is refined to increase the probabilities of interactions between vehicles, leading to more effective scenario generation.
\section{Conclusion}\label{sect:conc}

In this paper, we have proposed a novel fuzz testing method \tool, which addressed the limitations of existing methods and generated high-quality scenarios for autonomous driving systems. \tool combines a model-based fitness evaluation approach with distance-guided mutation strategies to improve the ability of fuzz testing in detecting violations. We have conducted extensive experiments to evaluate the effectiveness of \tool. The results show that compared to the state-of-art approaches AV-Fuzzer, DriveFuzz and TM-Fuzzer, \tool can detect 32, 27 and 9 more violations and demonstrate its ability to identify potential issues in the ADS.

In future work, we aim to further enhance scenario selection and mutation strategies. For example, we plan to incorporate factors such as traffic signals and weather conditions into the neural network model to improve the accuracy of fitness value evaluation. In addition, we intend to explore the use of sequence models such as Informer~\cite{zhou2021informer}, which has demonstrated superior performance over Transformers in extracting temporal features from trajectory prediction~\cite{10364872}, to enhance the effectiveness of the scenario violation prediction model. Furthermore, we aim to extend the application of \tool to other autonomous systems, such as unmanned aerial vehicles.


\bibliographystyle{ACM-Reference-Format}
\bibliography{software}

\end{document}